\documentclass[journal=jpccck,manuscript=article]{achemso}

\usepackage{hyperref}
\usepackage[latin1]{inputenc}
\usepackage{graphicx}
\usepackage{dcolumn}
\usepackage{bm}
\usepackage{ae}
\usepackage{aecompl}
\usepackage{amsmath}
\usepackage{amssymb}
\usepackage{epsfig}
\usepackage{psfrag}
\usepackage{color}
\usepackage[final]{pdfpages}

\renewcommand{\ell}{h}

\newcommand{\derpar}[3]{\left(\frac{\partial #1}{\partial #2}\right)_{#3}}

\author{{\L}ukasz Baran}
\affiliation{ Department of Theoretical Chemistry, Institute of Chemical Sciences, Faculty of Chemistry, Maria Curie-Sklodowska University in Lublin, Lublin, Poland.  }
\alsoaffiliation{Departamento de Qu\'{\i}mica F\'{\i}sica, Facultad de Ciencias
Qu\'{\i}micas, Universidad Complutense, Madrid, 28040, Spain.}
\email{lukasz.baran@mail.umcs.pl}
\author{Pablo Llombart}
\affiliation{Departamento de F{\'i}sica Te\'orica de la Materia Condensada, Instituto Nicol\'as Cabrera, Universidad Aut\'noma de Madrid, Madrid 28049, Spain}
\author{Luis G. MacDowell}
\affiliation{Departamento de Qu\'{\i}mica F\'{\i}sica, Facultad de Ciencias
Qu\'{\i}micas, Universidad Complutense, Madrid, 28040, Spain.}
\email{lgmac@quim.ucm.es}

\title{Understanding  Interfacial Ice Premelting:  Structure, Adhesion and Nucleation}


\begin{document}

\begin{abstract}
The interface of ice with solids plays an exceptionally important role on a wide variety of natural phenomena, such as the melting of permafrost, sliding of glaciers or frost heaving; as well as on many important technological applications such as windmills, car tyres or aircrafts. In this work, we perform a systematic computer simulation study of ice premelting, and explore the thickness and structure of quasi-liquid layers formed at the interface of ice with substrates of different hydrophilicity. Our study shows that interfacial premelting occurs systematically on neutral substrates of whatever hydrophilicity, forming films of limited thickness for substrates with contact angles larger than ca. 50$^\circ$ but exhibiting complete interfacial premelting at smaller contact angles.  Contrary to most experimental studies, we focus not only on the premelting behavior with temperature, but also with pressure, which is a matter of relevance in important situations such as  ice friction. Our study is guided within a rigorous framework of surface thermodynamics, which allows us to show that the premelting film structure is a function of a single thermodynamic variable. By this token we are able to relate properties measured along an isobar, with premelting films at arbitrary temperature and pressure. Our results are also exploited to study  ice adhesion, with a view to the understanding of icephobicity. We find that adhesion strength in atomically smooth surfaces is one to two orders of magnitude larger than those found in experiments, and conjecture that the reason is substrate roughness and the presence of organic adsorbents. Our theoretical framework also allows us to exploit our results on interfacial premelting in order to gain insight into heterogeneous ice nucleation. Our results show that apolar smooth substrates of whatever hydrophilicity are unlikely nucleators, and that too large hydrophilicity conspires also against ice nucleation. Furthermore, we exploit our statistical-thermodynamic framework to shed light into the nature of the surface intermolecular forces promoting interfacial premelting, and provide a  model to predict quasi-liquid layer thickness as a function of the substrate's hydrophilicity with great potential applications in fields ranging from earth sciences to aircraft engineering.
\end{abstract}




\section{1. Introduction}

The freezing of ice is tricky. In pure water, homogeneous nucleation of ice occurs at significant rates only for temperatures below -38$^\circ$~C.\cite{kanji17} Freezing at higher temperatures requires the action of heterogeneous nucleation, but the nucleation efficiency of most solids is limited and the underlying reasons for their nucleation ability are not fully understood.\cite{cantrell05,kiselev16,qiu17,kanji17,fitzner20,huang24}
Paradoxically, both water vapor and water droplets appear to freeze readily
on critical infrastructures such as aircrafts, windmills or solar panels, causing loss of efficiency, damage and hazards.\cite{jelle16,andersson17,irajizad19} For this reason, much effort is devoted to the design of ice-phobic surfaces able to limit the accretion of ice and facilitate its removal.\cite{li24,pradhan24} 

A fundamental property to characterize ice-phobicity is ice adhesion.\cite{emelyanenko20,stendardo24,rehfeld24} A noticeable feature of adhesion tests is that the tensile adhesion is always larger than the shear adhesion.\cite{jellinek62} This observation together with the well-known low friction coefficient of ice was considered already in the 1950s as evidence for the existence of interfacial ice  premelting, i.e: the formation of a so called {\em quasi-liquid layer} of molten ice in between the bulk ice phase and the substrate at temperatures below the melting point.\cite{jellinek62,jellinek67,dash06}

For some time, this was a controversial hypothesis (c.f. \cite{weyl51} for an early account), but strong evidence has been gathered since the 1980s showing the ubiquity of interfacial premelting of ice, both from experiments,\cite{maruyama92,gay92,furukawa93,ishizaki96,pittenger01,engemann04,schoder09,li19} theory,\cite{fletcher62,lipowsky90} and computer simulations.\cite{nikifioridis21, baran22, sun22, zhu23, uchida23,cui24} Indeed with very few exceptions,\cite{beaglehole94} experiments of ice premelting in different materials, ranging from graphene, to glass, clay, or other minerals have systematically shown evidence of the formation of a quasi-liquid layer of significant thickness.\cite{maruyama92,gay92,furukawa93,ishizaki96,pittenger01,engemann04,schoder09,li19}

The understanding of ice premelting is a topic of great interest per-se, as it plays a fundamental role in important natural phenomena, such as regelation, glacier sliding, frost heaving, and soil stability;\cite{dash95,dash06,gerber22,wan22,style23,meyer24} as well as transportation safety in cold regions, skiing and skating.\cite{colbeck88,lozowski13,lever21b,du23b} However, understanding premelting can also provide complementary understanding of substrates ability to act as a heterogenous driver of ice formation, with implications in atmospheric sciences,\cite{bartels12} icephobic materials, and ice adhesion.\cite{emelyanenko20} 

The theory of wetting provides a unifying framework in support of this complementarity.\cite{schick90} Indeed the nucleation activity can be quantified in terms of the spreading coefficient of ice in between substrate and water.\cite{marks23,vicars24} On the contrary, the study of interfacial melting is related to the spreading coefficient of a water film in between ice and a substrate.\cite{emelyanenko20} Therefore, we expect that the study of ice premelting, which is far less often considered than heterogeneous nucleation can provide very interesting and complementary insight into the problem. A particularly important question is whether we can expect any nucleation ability from a substrate which exhibits propensity for ice premelting.\cite{uchida21,zhu23}

In this paper, we perform computer simulations of ice in contact with substrates of tunable hydrophilicity. Our work shows that for a family of simple face centered cubic substrates, interfacial premelting appears to occur spontaneously for substrates with contact angles greater than 50$^\circ$. Our results can be rationalized under the framework of wetting physics and we show that the properties of premelting films, both thickness and structure, can be described in terms of one single thermodynamic variable. By this device we are able to infer properties of interfaces measured along an isobar with those of interfaces measured along and isotherm. Our analysis also gives us great insight into the nature of surface intermolecular forces governing ice pre-melting and show that the popular model of dispersion forces used often for theoretical studies can only account for a very small fraction of the full interactions. Our results are also exploited to characterize adhesion and nucleation ability as a function of hydrophilicity.

\section{2. Thermodynamics of interfacially premelted films}

\label{sec:theory}

The thermodynamics of interfacial premelting is usually described in terms of
the Gibbs free energy.\cite{dash95,wettlaufer06,style23} Whereas such approach gives correct results, it
rises some conceptual problems.\cite{style23} The reason is that the Gibbs free energy is best
formulated for systems of constant temperature and pressure.\cite{callen85} However, the
pressure of inhomogeneous systems is not a scalar, but a tensor, and can adopt
different values in different parts of the system.\cite{henderson92b}

For this reason, it is more cautious to cast the problem in terms of the grand free energy, i.e., the thermodynamic potential akin to systems at constant temperature and chemical potential. The advantage is that the chemical potential is a scalar, and the condition of material equilibrium requires it to be equal everywhere in the system. As a result, minimizing the grand potential subject to the constraint of fixed chemical potential of a bulk reservoir is a safe procedure.\cite{henderson05} 

This conceptual advantage has been widely recognized in the physics of wetting,
where the problem is consistently formulated under the constraint of the
constant chemical potential imposed by the bulk vapor phase in equilibrium with
a substrate.\cite{derjaguin87,dietrich88,schick90,henderson92b,evans92} In the section below, we extend these ideas to the problem of
interfacial melting, and show that the powerful machinery of wetting physics
provides a rigorous and consistent framework for the description of interfacial
premelting too.

\subsection{2.1. Theoretical formulation}

In the important case that ice is in contact with an inert
substrate, the study of interfacial premelting seeks to characterize how a
microscopic quasi-liquid layer of melt-water intrudes at the interface of ice
with the inert substrate. However, for the sake of generality, in this section we will formulate the problem for the general case where a liquid phase, $\beta$, intrudes at the interface between some inert substrate $S$ and a solid phase $\alpha$ close to its melting point.
 
Consider a bulk solid phase, $\alpha$, in equilibrium with an inert solid substrate,
$S$. In the interfacial region where both phases meet, the  sharp
termination of the solid has a strong enthalpic cost as the interfacial
atoms loose about half of their cohesive energy.\cite{israelachvili11}  In order to relax the
enthalpic penalty, the interfacial region of the solid phase can undergo
significant structural changes which help decrease the overall free energy
budget.  
If phase $\alpha$ is well inside its region of stability, and the temperature is
low, the changes are likely to be small  rearrangements of the lattice structure which help
optimize the enthalpy of interaction between the solid and the substrate.  
However, if the the temperature is sufficiently high, the minimization of the
interfacial free energy can also take place by increasing the entropy of
the interfacial region. Such a mechanism is particularly likely to happen when
the solid phase is close to its melting line. In this situation,
the system can bare the melting of finite amounts of bulk liquid phase, $\beta$,
in order to increase the interfacial entropy, and correspondingly, to decrease
the overall surface free energy.

In order to account for the total free energy budget, it is convenient to
consider a system at fixed temperature, $T$ and chemical potential, $\mu$. Under such
constraints, the relevant thermodynamic potential is the grand free energy,
$\Omega(T,\mu)$, which can be readily obtained from the Helmholtz free energy by 
a Legendre transformation $A \to A - \mu n$, where $n$ is the number of molecules.

In the bulk, this potential is particularly simple, because it amounts to $\Omega = - pV$, 
where $p=p(T,\mu)$ is the bulk pressure and $V$ is the system's volume. At phase coexistence,
both the liquid and the solid have equal pressure. Therefore, the melting of
a small amount $\Delta V$ of the solid phase leads to a change $-\Delta V$ of
the liquid phase. i.e. since the pressure in both phases is equal, the transformation of one 
phase into the other occurs with no change in the total grand free energy of the system. 

It follows that exactly at coexistence the only relevant contribution to the system is the surface free energy, which, will depend variationally on the amount of
liquid formed next to the wall, as measured by the quasi-liquid layer thickness
$h$. 

As a reference, we first consider the special
case of complete surface melting. This is the situation that occurs when the system is exactly at solid/liquid coexistence, and the
compatibility of the substrate with the liquid phase is so much larger than with
the solid phase, that a full bulk liquid phase is formed in between. As a result, the  total surface free energy  
$\omega(h,T,\mu)$, is the cost of forming independent substrate/liquid and liquid/solid interfaces:
\begin{equation}\label{eq:surfmelt}
    \omega(h\to\infty;T,\mu) = \gamma_{S\beta} + \gamma_{\alpha\beta}
\end{equation} 
If, on the other hand, we move into the region of stability of the solid phase,
the bulk liquid, $p_{\beta}(T,\mu)$ and bulk solid pressures,
$p_{\alpha}(T,\mu)$ at the prevailing temperature and chemical potential, are no longer equal. 
The conversion of solid into liquid therefore is now penalized by a bulk free energy term
$-(p_{\beta}-p_{\alpha})A h$, with $A$ the lateral surface, so that a bulk liquid phase of infinite
extent cannot possibly form. 

However, we still must acknowledge that the
system could have a propensity to form at least a finite amount of liquid phase
in between the substrate and the solid. In surface thermodynamics, such a
propensity is accounted by means of the interface potential, $g(h)$ which
measures the free energy per unit surface of a premelting film at coexistence as a function
of the premelting film thickness $h$.\cite{schick90,dietrich88,churaev88,henderson05,starov09,emelyanenko20} 

With the help of the interface potential, the free energy balance
in the general case of a system away from coexistence is given by:
\begin{equation}\label{eq:fullfree}
   \omega(h;T,\mu) =  \gamma_{S\beta} + \gamma_{\alpha\beta} + g(h) - \Delta p h
\end{equation} 
where $\Delta p(T,\mu) =  p_{\beta}-p_{\alpha}$ is the pressure difference between
bulk liquid and solid at the imposed chemical potential. 
Notice that here, $g(h)$ accounts for the effective interaction between the wall/liquid and liquid/solid interfaces, while
$\Delta p h$ accounts for the cost of forming a bulk liquid at conditions away
from the melting line. Also notice that in view of the conventions employed, $g(h)$ must vanish in the limit
that $h\to\infty$ in order for Eq.(\ref{eq:fullfree}) to recover the expected
result of Eq.(\ref{eq:surfmelt}) for complete surface melting.

Our aim here is to determine what is the equilibrium film thickness, $h_e$ at the
imposed temperature and chemical potential, given that we know $g(h)$. This is
found by requiring $\omega(h;T,\mu)$ to become minimum, i.e., such that
$d\omega/dh = 0$, which leads to the condition:
\begin{equation}\label{eq:eqcond}
          \Pi(h_e) = -\Delta p
\end{equation} 
where, by definition, $\Pi(h)=-dg/dh$ is known as the disjoining pressure.\cite{derjaguin87}

Once $h_e$ is known, the equilibrium surface free energy of the system can be calculated
by substitution of $h_e$ into Eq.(\ref{eq:fullfree}). Particularly, for a system exactly at
the coexistence chemical potential $\mu_c$, this yields:
\begin{equation}\label{eq:omegaeq}
     \omega_e(T,\mu_c) =  \gamma_{S\beta} + \gamma_{\alpha\beta} + g(h_e) 
\end{equation} 
Since this measure of the free energy corresponds exactly to the equilibrium of
the substrate with the solid phase, it must therefore correspond to the 
substrate/solid surface tension, such that  $\omega_e(T,\mu_c)=\gamma_{S\alpha}$. Therefore, the above
equation acknowledges explicitly that the {\em equilibrium} interfacial tension of a solid
phase with a substrate can exhibit a finite thickness liquid film, a fact that
is not always appreciated. Replacing the above identity into
Eq.(\ref{eq:omegaeq}), leads to an important constraint on the value of
$g(h_e)$, i.e.:
\begin{equation}\label{eq:gmin1}
   g(h_e) = \gamma_{S\alpha} -  \gamma_{S\beta} - \gamma_{\alpha\beta}
\end{equation} 
which, by virtue of Young's equation for a liquid droplet forming in between 
the substrate and the solid phase, may be also expressed as:\cite{churaev88,starov09}
\begin{equation}\label{eq:gmin2}
  g(h_e) = \gamma_{\alpha\beta} ( cos(\theta) - 1 )
\end{equation} 
where $\theta$ is the droplet's contact angle.

To show the consistency of these results, consider the special case of complete
surface melting, which, in terms of the film thickness implies that
$h_e\to\infty$, and accordingly, that  $g(h_e)\to 0$. Replacing this result
into Eq.(\ref{eq:gmin1}) leads to the condition for complete wetting of the substrate
by the phase $\beta$, while replacing it into  Eq.(\ref{eq:gmin2}), it implies
$\theta=0$ as expected.

Unfortunately, our theoretical knowledge is usually not sufficient to fully
constrain $g(h)$, so that  Eq.(\ref{eq:eqcond}) is not  a means of predicting
the equilibrium film thickness, but instead, an operational procedure to evaluate
the first derivative of $g(h)$ from knowledge of $\Delta p$.\cite{blake72} 

Despite of this shortcoming,  Eq.(\ref{eq:eqcond}) is still a useful starting
point for the characterization of quasi-liquid layers. The reason is that
$g(h)$ is only weakly temperature dependent. Therefore, to a first approximation
 Eq.(\ref{eq:eqcond}) implies that $h$ could be formally retrieved as $h=\Pi^{-1}(\Delta p)$, 
where $\Pi^{-1}(\Delta p)$ is a function of the single variable  $\Delta
p(T,\mu)$ (c.f. Ref.~\cite{henson04}). This is convenient, because $h$ is an interfacial property, but
$\Delta p(T,\mu)$ is a purely bulk property that can be calculated by
thermodynamic integration. i.e., 
if we somehow measure $h$ for a given set of state points
in the two dimensional space of the phase diagram, we can predict $h$ for any
other state provided we measure its distance away
from the melting line in terms of the single variable  $\Delta p(T,\mu)$.

\begin{figure}
\begin{center}
\includegraphics[width=0.45\textwidth]{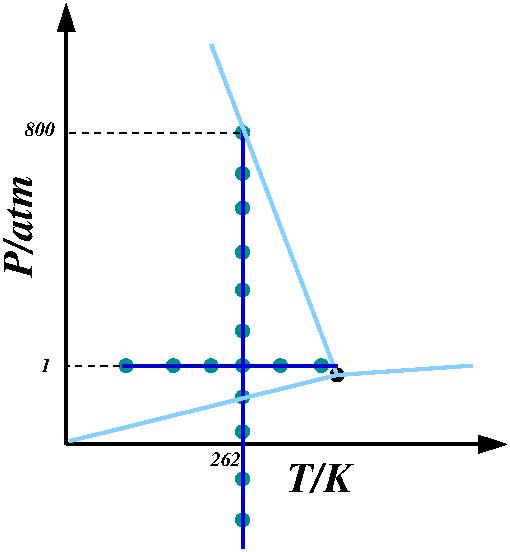} 
\end{center}
\caption{\label{fig:coexistence}
Schematic phase diagram of water in the neighborhood of the triple point.
The lines display schematically the isotherm and isobars simulated in this study.
}
\end{figure}

\subsection{2.2. Calculation of $\Delta p$}

Of course, $\mu$ is not a convenient thermodynamic variable to describe a state
point in the phase diagram. Here, we are assuming that the system contains
infinite amounts of the solid phase, so that it acts as a reservoir fixing
the chemical potential for a fixed volume subsystem next to the substrate. Accordingly, in what follows we will describe the thermodynamic state
of the system from knowledge of $T$ and $p_\alpha$, i.e. the system's
temperature and the perpendicular pressure the solid bulk phase exerts on the wall. Notice we are still working under the constraint of constant
temperature and chemical potential of the system, but we take into account that
$\mu=\mu(T,p_\alpha)$, so that we can characterize each state point $(T,\mu)$ in
terms of $(T,p_\alpha)$.

Having acknowledged this, a programme for the characterization of the premelting film thickness may be
accomplished by measuring $h$ for different pressures of the solid phase along an isotherm,
then estimating $\Delta p(T,p_{\alpha})$ for each state point by thermodynamic
integration, and finally using the resulting function to predict $h$ at arbitrary points
in the phase diagram.

For the calculation of  $\Delta p(T,p_{\alpha})$, we consider the thermodynamic
relation for the chemical potential of an arbitrary single component substance:
\begin{eqnarray}\label{eq:dmu}
  d\mu(T,p) = -s dT + v dp
\end{eqnarray}
where $s$ and $v$ are the molar entropy and volume, respectively.

In order to exploit this equation, we choose an arbitrary point lying along
the melting line, where the solid and liquid phases have exactly the same
chemical potential, and then allow the temperature and pressure of each phase
to change arbitrarily under the constraint that both phases remain in
states of equal chemical potential. This can only be achieved by requiring
that the increments of chemical potential remain equal, 
$d\mu_{\alpha}(T_{\alpha},p_{\alpha})= d\mu_{\beta}(T_{\beta},p_{\beta})$,
so that:
\begin{equation}\label{eq:cstmu}
     -s_{\alpha} dT_{\alpha} + v_{\alpha} dp_{\alpha} =
     -s_{\beta} dT_{\beta} + v_{\beta} dp_{\beta}
\end{equation}

\subsubsection{Path along an isotherm}

For isothermal paths at temperature $T$, we change the
pressure of the solid phase ($\alpha$) from the melting pressure $p_m(T)$ to
an arbitrary value of pressure $p_{\alpha}$, and seek for the corresponding
value of pressure in the liquid phase ($\beta$). In this case of course $dT=0$, and
the above equation can be solved formally as:
\begin{equation}
  \int_{p_m}^{p_{\alpha}} v_{\alpha}(T,p') dp' = \int_{p_m}^{p_{\beta}}
v_{\beta}(T,p'') dp''
\end{equation}
For a first approximation, we can consider that the liquid and solid phases
are incompressible, so that the volume of each phase may be assumed
 equal to the value adopted at phase coexistence. Accordingly, we find:
 \begin{equation}\label{eq:dpisotherm}
   p_{\beta} - p_{\alpha} = \left (\frac{v_{\alpha}}{v_{\beta} }-1\right)(p_{\alpha} - p_m(T))
\end{equation}
This equation remains qualitatively correct for a wide range of pressures, but
unfortunately, is not quantitatively correct for large deviations of
$p_{\alpha}$ away from the melting pressure.\cite{style23,zhu23,baran24b} To improve the accuracy,
we assume a linear dependence  of molar volume on pressure,
$v(T,p) = A + B p$, and integrate to obtain:\cite{baran24b}
\begin{equation}
    A_{\beta}(p_{\beta}-p_m) + \frac{1}{2}B_{\beta}(p_{\beta}^2 - p_m^2) = A_{\alpha} ( p_{\alpha} - p_m) +
\frac{1}{2} B_{\alpha}(p_{\alpha}^2 - p_m^2)
\end{equation}
This is a quadratic equation that can be readily solved for $p_{\beta}$ as a
function of the control variables $T$ and $p_{\alpha}$. The volume coefficients
may be calculated from a linear fit to pVT data obtained from simulation.

\subsubsection{Path along an isobar}

An alternative route that is experimentally more convenient is to fix the
pressure of the solid in contact with the substrate and perform measurements
of the film thickness by changes in the system's temperature.

In this case, we use Eq.(\ref{eq:cstmu}), but take into account that now
$dp_{\alpha}=0$. This leads to: 
\begin{equation} \label{eq:cstmup}
   dp_{\beta} = \frac{\Delta s(T,\mu)}{v_{\beta}(T,\mu)} dT
\end{equation}
where $\Delta s(T,\mu)=s_{\beta}(T,p_{\beta}(T,\mu)) -
s_{\alpha}(T,p_{\alpha}(T,\mu))$.

For small deviations of the temperature away from the melting point, we
may assume the entropy difference along the isobar is equal to
that at the same temperature along the melting line, such that
$\Delta s= \Delta h_m/T$, with
$\Delta h_m=h_{\beta}-h_{\alpha}$ the heat of melting.
If we also assume the liquid
phase is incompressible, and perform an integration from an arbitrary
temperature $T$ to the
melting temperature $T_m(p_{\alpha})$, we obtain:
\begin{equation}\label{eq:dpisobar}
   p_{\beta} - p_{\alpha} = \frac{\Delta h_m}{v_{\beta}}
      \ln\frac{T}{T_m(p_{\alpha})}
\end{equation}

Unfortunately, this equation is accurate only for deviations of about 10~K away
from the melting temperature. For larger undercooling, it is required to account
for the entropy and volume change along the thermodynamic path. Particularly for
the entropy we need to invoke the standard thermodynamic relation:
\begin{equation}\label{eq:entropy}
  dS = \derpar{H}{T}{p} \frac{dT}{T} - \derpar{V}{T}{p} dp
\end{equation}
which has to be integrated along with Eq.(\ref{eq:cstmup}) for both the
liquid and the solid phase. 
In practice, this can be implemented by an on the fly integration procedure
which couples differential equation solvers with computer simulations required
for the determination of the thermal coefficients, similar to \cite{baran24b}
(see Methods section).

\subsubsection{Consistency check}

Notice that the pressure difference $p_{\beta}-p_{\alpha}$ is a purely
bulk property which does not depend on the  path of integration. Whence,
Eq.\ref{eq:dpisotherm} and Eq.\ref{eq:dpisobar} provide the same pressure difference, but expressed in
terms of different measures of distance away from the melting line. i.e,
$p_{\alpha}-p_m(T)$, for Eq.\ref{eq:dpisotherm} or $T/T_m(p_{\alpha})$ for
Eq.\ref{eq:dpisobar}.

To check this, we invoke the Clausius equation for phase coexistence,
$\frac{dp}{dT}=\frac{\Delta h_m}{\Delta v T}$. For the  special case of
coexistence between two incompressible phases, this
result can be readily integrated from a sought point along the
melting line $(T,p(T))$ to some other known point along that
line $(T^*,p^*)$ (e.g.: the triple point, or the normal melting point), 
yielding:
\begin{equation}
   p(T) - p^{*} = \frac{\Delta h_m}{\Delta v} \ln\frac{T}{T^{*}}
\end{equation} 
Using this result to replace $p_m(T)$ in Eq.\ref{eq:dpisotherm} or to eliminate 
$T_m(p_{\alpha})$ in Eq.\ref{eq:dpisobar} leads in both cases to:
\begin{equation}
   p_{\beta}-p_{\alpha} = (1-\frac{v_{\beta}}{v_{\alpha}}) (p^* - p_{\alpha}) +
	   \frac{\Delta h_m}{v_{\beta}}\ln\frac{T}{T^*}
\end{equation} 
Where now the pressure difference is expressed explicitly in terms of the
control parameters $T,p_{\alpha}$, and a reference point $(T^*,p^*)$ along the
melting line.

The above expression is a non-linearized version of the result often known
as the {\em Generalized Clapeyron equation} in the study of premelting dynamics and regelation \cite{wettlaufer06,style23}.

\section{3. Model and Computer Simulations}

\paragraph*{Model}
The simulations are carried out using the TIP4P/Ice model for water,\cite{abascal05} albeit with Lennard-Jones interactions truncated at 1~nanometer. The wall is made as a stack of face centered cubic (FCC) atoms completely fixed in their lattice positions. Wall atoms interact  via Lennard-Jones (LJ) interactions with water oxygens only, with wall-oxygen interactions truncated at 1~nm also. 
The wall Lennard-Jones parameters are set such that $\epsilon_{wall}=f\epsilon_{O}$, and $\sigma_{wall}=\sigma_{O}$, where the sub-indexes $wall$ and $O$, stand for the wall and TIP4P/Ice oxygen atoms, respectively. The hydrophilicity of the wall is tuned by changing $f$. Simulations have been carried out for $f=1, 2, 3$ and $4$, corresponding to water contact angles of $\theta=120^\circ, 107^\circ, 91^\circ$ and $50^\circ$ at 
$T=298$~K, respectively.\cite{baran22}


\paragraph*{Ice:} An initial configuration of hexagonal ice (Ih) is prepared from a pseudo-orthorombic unit cell  comprising of 8 molecules with lattice parameters $a$, $b=\sqrt{3}\,a$ and $c$, aligned such that the $c$ axis lies along the hexagonal $c$ axis.\cite{macdowell10} The full system contains a stack of  $8\times4\times15$ such unit cells, with a total of  3840 water molecules.  An initial hydrogen bond network following the Bernal-Fowler rules is prepared using the method of Ref.\cite{buch98}. This configuration is then set to a total zero dipole moment by performing a Monte Carlo simulation that accepts hydrogen bond network arrangements that decrease the dipole moment.\cite{moreira18} To sample the hydrogen bonds while obeying the Bernal-Fowler rules, we use a ring rotation algorithm.\cite{rick03,macdowell10} 

\paragraph*{Wall:} The model FCC lattice is arranged along its (111) direction, with the unit cell dimensions chosen so as to be fully congruent with the basal face of the ice lattice. This can be achieved by generating a FCC pseudo-orthorombic unit cell with sides $a'=a$, $b'=\sqrt{3}\,a$ and $c'=\sqrt{6}\,a$. Given this convention, the fractional coordinates of the wall atoms in the pseudo-orthorombic unit cell are $(0,0,0)$, $(1/2,1/2,0)$, $(0, 1/3, 1/3)$, $(1/2, 5/6, 1/3)$ $(0,2/3,2/3)$ and $(1/2, 1/6, 2/3)$. 

\paragraph*{Sample preparation}

A compound wall-ice system is prepared by aligning the $c$ axis of ice and the $c'$ axis of the wall along the $z$ direction of the simulation box. This leaves the (111) plane of the FCC wall parallel to the (0001) basal plane of ice. 

For each temperature and pressure studied, the unit cell dimensions of ice are calculated from bulk NPT simulations.  From this data, the ice slab is re-scaled to the bulk dimensions of the corresponding thermodynamic state before the start of the equilibration period. The dimensions of the wall are scaled accordingly. The ice stack in the $z$ direction is prepared such that it leaves half a bilayer exposed at the  surfaces on each side of the slab. This serves to promote premelting when the slab is brought into contact with the wall.
Once the initial configuration is so prepared, the ice slab is  gently sandwiched against two walls to the desired pressure, and the resulting premelting film is allowed to equilibrate for at least 60 nanoseconds. The sandwiching and thermalization are accomplished using a {\em wall barostat}.\cite{heyes11,marchio18,amabili19,baran22}  This imposes a fixed target pressure along the direction perpendicular to the interface and together with the temperature fixes completely the thermodynamic state of the inhomogeneous wall-ice system to $T,\mu(p_{\perp})$.\cite{macdowell07} The lateral dimensions of the simulation box are fixed and interact via periodic boundary conditions along the $x$ and $y$ directions. In practice, periodic boundary conditions on the $z$ direction are also applied, but  a distance of 4~nm is left between the walls to avoid spurious interactions of the interface. The whole system is made large enough that the parallel pressure in the middle of the simulation box has become equal to the imposed perpendicular pressure. This corresponds to the limit of an infinite ice slab in contact with a wall and removes effects related to the finite size of the simulation box in the $z$ direction (notice that for finite size systems the tangential pressure $p_\parallel$ need not equal $p_{\perp}$, and the intensive properties of the system then become a function of three thermodynamic variables, ($T,\mu,p_{\perp}$)).\cite{evans87,henderson92b,macdowell07}  

\paragraph*{Wall barostat}

The thermodynamic state of the system is enforced by using a wall barostat, which effectively achieves a $Np_{\perp}AT$ ensemble in a natural way.\cite{heyes11,marchio18,amabili19,baran22} This is done by imposing a constant uniform force $F_z=p_{\perp}A/N_w$ directed towards the ice sample to each of the $N_{wall}$ wall atoms (with $A$ the lateral area of the system). It is not possible to use the virial formula for the pressure to check the consistency of this barostat, because the volume in this inhomogeneous system is not well defined. Therefore, one must resort to the method of planes, which allows to measure explicitly the force across a plane, and only requires knowledge of forces and total area of the system to provide the correct pressure.\cite{todd95} For practical matters, this corresponds to  checking  that the average force exerted by the wall atoms on the water molecules during the simulation corresponds exactly to the target pressure.\cite{henderson92b,baran22}

\paragraph*{Molecular Dynamics Simulations}

All of the simulations have been launched using the  LAMMPS simulation package \cite{lammps22}. 
Trajectories are evolved with the velocity-Verlet algorithm, using a time step of $2$~fs. Bonds and bond angles are constrained by the use of SHAKE algorithm. 
The system is thermostated by the use of a velocity rescale algorithm with damping factor $\tau_{VR}=2$~ps.\cite{bussi07} In the auxiliary $NpT$ simulations, both the temperature and pressure were maintained using a three chains Nos{\'e}-Hoover algorithm with  a damping factor $\tau_{NH}=2$~ps. Lennard-Jones interactions were truncated at a distance of $10$~\AA~for water-water interactions. Tail correction were not included. Long-range electrostatics were calculated using  the particle-particle particle-mesh (pppm) method.\cite{darden93}  The charge structure factors were evaluated with a grid spacing of $1$ \AA~and the fourth order interpolation scheme.
It resulted in the $36 \times 32 \times 120$ vectors in the $x, y, z$ reciprocal directions for simulations of ice sandwiched between two walls. In the case of simulations performed to evaluate the melting pressure at $T=262$~K,  it resulted in  $144 \times 64 \times 60$ and $72 \times 64 \times 120$ vectors in the $x, y, z$ reciprocal directions for pII and basal planes, respectively. 

\paragraph*{Order parameter and film thickness}

To distinguish between the solid- and liquid-like environments 
we used the CHILL+ parameter.\cite{nguyen15} Based on the calculation of
correlation function $c_3(i,j)$ which takes on different values
for staggered and eclipsed conformations, this allowed us to discriminate
between different allotropic forms of ice. Molecules labelled
as ice Ih and Ic, including interfacial Ih molecules were assigned
as solid-like, whereas the remaining types,
 including mislabelled molecules were assigned to the liquid phase. 
 Having the average number of liquid water molecules, we have determined the equilibrium film height as $0.5\times N_{liq}/(A\times\rho_{liq}(T,p))$, with due account of the density variation of liquid molecules with temperature and pressure. Also notice the $1/2$ prefactor takes into account the presence of two interfaces in our simulation setup.

\paragraph*{Melting Pressure}

To determine the melting pressure of ice Ih at $T=262$ K we have employed a direct coexistence method.\cite{garcia06} This method brings explicitly two phases at coexistence across an interface. Two systems with different interfaces were studied, namely, (i) the basal (0001) and (ii) the secondary prism ($\bar{1}2\bar{1}0$) faces. The simulations were performed
at different pressures in an $Np_{\perp}AT$ ensemble to estimate the melting/freezing rates based on the time evolution of number  of liquid molecules, as identified using the CHILL+ algorithm.\cite{nguyen15} The melting/freezing rates were plotted as a function of pressure, and the melting pressure was located by interpolation of results to zero melting/freezing rate.  The simulations were performed in an Np$_{\perp}$AT ensemble using  a large system of $15360$ molecules. This allows for minimization of large scale random fluctuations that seriously hamper the location of melting points.\cite{conde17} Despite the choice of large systems, the  simulations required about $40-80$ ns to safely determine the average melting/freezing rates (c.f. Supplementary Material). \\

\paragraph*{Thermodynamic integration}

The calculation of the pressure difference between the solid $(i)$
and liquid $(w)$ phases, required for the evaluation of disjoining pressure curve, is performed by solving the system of two coupled
differential equations Eq.~\ref{eq:cstmup} and Eq.~\ref{eq:entropy}
as in the Gibbs-Duhem integration.\cite{kofke93} 

At each state ($T', p'$) the water's pressure $p_w(T')$
is calculated as 

\begin{equation}
    p_w(T')=p_w(T)+
    \frac{\Delta S(T)}{V_w(T,p_w)}(T'-T)
    \label{eq:gibbs-duhem1}
\end{equation}

\noindent  The entropy change $\Delta S$, is estimated as

\begin{equation}
\Delta S(T')=\Delta S(T)+\Delta C_{p}\ln\frac{T'}{T}-V_{w}(p'-p)
\label{eq:gibbs-duhem2}
\end{equation}

For ice, thermodynamic quantities are calculated at the given temperature and a constant isobar pressure of $p_i=1$~atm, whereas the properties of 
liquid water are calculated at the same temperature and
the  liquid pressure $p_w(T)$, as estimated from Eq.\ref{eq:gibbs-duhem1}.

The procedure starts at the assumed melting point,  
$T=T_m=270$~K for $p_i=p_w=p_m=1$~atm, where $\Delta S=\Delta H_m/T_m$ holds exactly. The initial conditions for the integration are provided in  Table~\ref{therm}. Notice the results differ slightly from the canonical TIP4P/Ice model (c.f. the Table IV in Ref.~\cite{abascal05}), as our simulations have the Lennard-Jones interactions fully truncated at a cutoff distance of 1~nm, with no tail corrections added. 
\begin{table}[h!]
\centering
\begin{tabular} {ccccc}
\hline
\hline
$V_w$ (m$^3\cdot$ mol$^{-1}$) & $\Delta H_m$ (J$\cdot$ mol$^{-1}$)  &$\Delta S_m$ (J$\cdot$ K$^{-1}\cdot$ mol$^{-1}$) & $T_m$ (K) & $p_i$ (atm)\\
\hline
$1.85\cdot10^{-5}$ &$5049.376$ & $18.701$ & 270 & 1 
\end{tabular}
\caption{Thermophysical properties of TIP4P/Ice model required for the determination of the
water pressure along the line of equal chemical potentials of the bulk (ice) and 
adsorbed (water) phases at melting point under ambient pressure}
\label{therm}
\end{table}

For a given temperature $T'=T+\delta T$, with an increment 
of $\delta T=10 K$, we find a new
$p_w(T')$ from the equation \ref{eq:gibbs-duhem1}. 
Afterwards, we perform a new simulation at $T',~p_i=1$~atm
for the solid and at $T', p_w$ for the liquid
to estimate the derivative properties as:

\begin{align}
 C_p(T)=\left(\frac{\partial H}{\partial T}\right)_p&=\frac{H(T+\delta T,p)-H(T,p)}{\delta T} \nonumber \\
 V_{w}=\left(\frac{\partial V}{\partial T}\right)_p&=\frac{V(T+\delta T,p)-V(T,p)}{\delta T}
 \label{eq:derivatives}
\end{align}

These results are used to update $\Delta S(T')$ 
according to the Equation~\ref{eq:gibbs-duhem2}.
Afterwards, the procedure starts again 
with simulations at $T'$ and $p_i=1$~atm
for the solid phase and simulations of the liquid
at $T'$ and $p_w$ as the new reference points. 

Once the full curve of $p_w $ is known, the
disjoining pressure is evaluated for a given premelting
film height $h(T)$ as:

\begin{equation}
\Pi(h(T))=-\Delta p=-(p_w-p_i)    
\end{equation}

\section{4. Results and Discussion}

We have performed molecular dynamics simulations of ice in contact with a family of substrates of different wettability.   Water molecules are modeled with the TIP4P/Ice potential,\cite{abascal05} while the substrate mimicks an idealized crystalline solid with a face centered cubic arrangement of Lennard-Jones atoms. Wall atoms have the $\sigma$ Lennard-Jones parameter equal to that of oxygens, but we allow for a range of different wettabilities by tuning the {\em wall strength}, $f$. This is achieved by setting the Lennard-Jones energy between wall atoms and oxygens, $\epsilon_{wall}$ to $f$ times the energy between two oxygen atoms, $\epsilon_{O}$, such that $\epsilon_{wall}=f\epsilon_{O}$. We have explored wall strengths of $f=1,2,3,4$, which correspond approximately to contact angles of $\theta=120^{\circ},107^{\circ}, 91^{\circ}$ and $50^{\circ}$, respectively, thus changing the wettability from strongly hydrophobic to moderately hydrophilic substrates.\cite{baran22} 

In order to scan the premelting behaviour over a large set of thermodynamic states, we monitor the temperature of the sample, together with the perpendicular pressure exerted by the wall on the ice sample (c.f. Methods). We have checked that for a box sufficiently large in the $z$ direction, the wall barostat sets the density as expected for the bulk phase at the imposed temperature and pressure (c.f. Supplementary Material). This guarantees our simulation setup is effectively simulating an infinite large ice sample in contact with a wall.

Simulations have been performed mainly for two different batches. A batch along an isobar $p=1$~atm for temperatures ranging between 230 to 269~K, very close to the expected  normal melting point of TIP4P/Ice.\cite{conde17} Another batch is performed for the isotherm $T=262$~K, for pressures ranging from 900~atm, well above the estimated melting pressure $p_m=802$~atm of the model, to highly overstretched states of up to $p=-2000$~atm (c.f. Figure~\ref{fig:coexistence}).

\subsection{4.1. Structure of interfacially premelted films}

\label{sec:structure}

In order to characterize the extent of interfacial premelting, we use the CHILL+ order parameter, which allows one to distinguish whether a single water molecule is surrounded by a liquid like or solid like environment.\cite{nguyen15}

\begin{figure}
\begin{center}
   \includegraphics[width=0.85\textwidth]{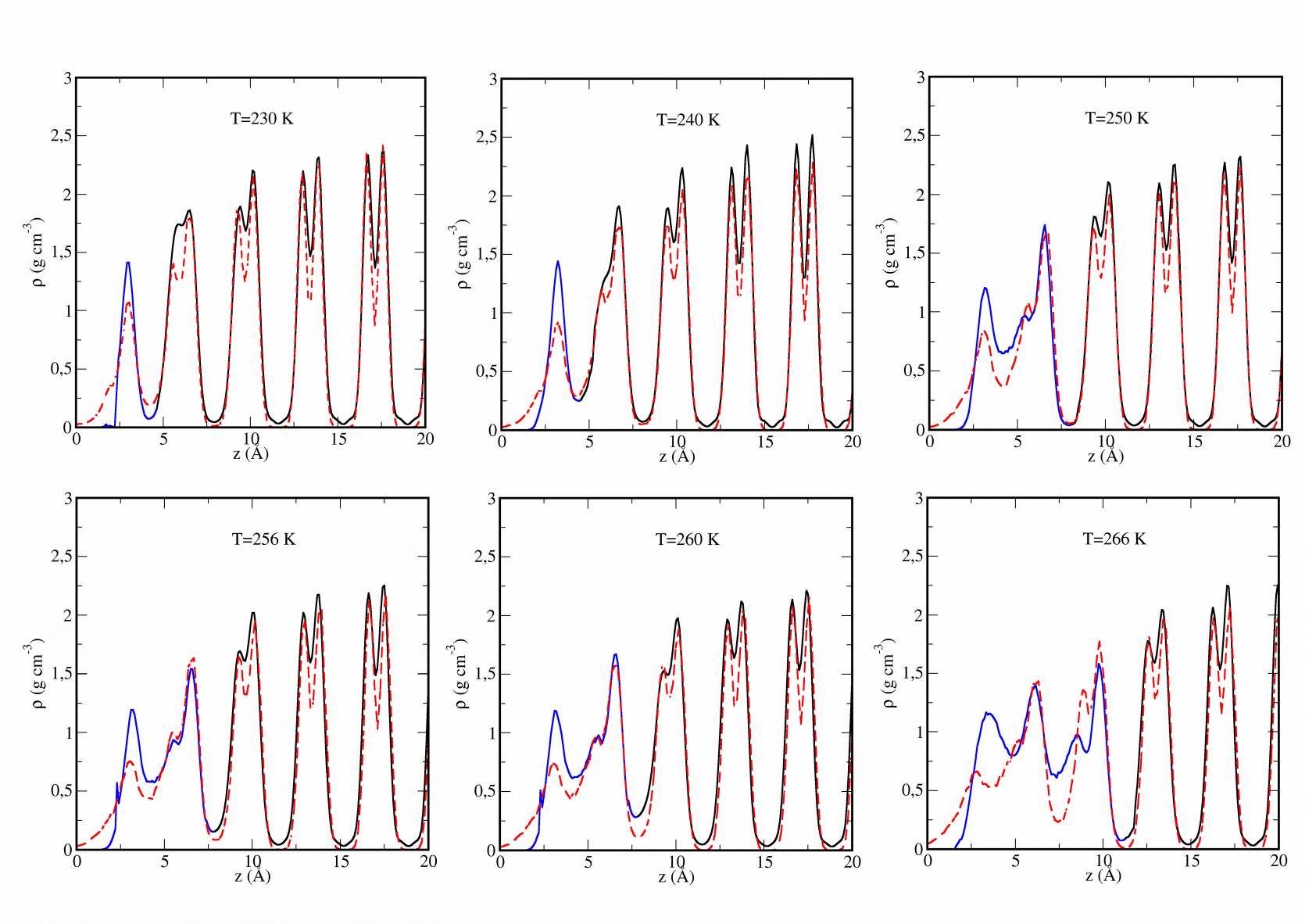} 
\end{center}
\caption{\label{fig:density_p1_f1}
   Density profiles along the  $p=1$~atm isobar  for the hydrophobic system with $\theta=120^\circ / f=1$. The continuous line is divided in two,
   with blue color depicting the region where liquid-like water is the majority phase and black color where ice is the majority phase. 
   The red dashed line corresponds to the ice-vapor density profile
   at the same thermodynamic conditions as obtained from Ref.\cite{llombart19,llombart20}. 
}
\end{figure}

\begin{figure}
\begin{center}
   \includegraphics[width=0.85\textwidth]{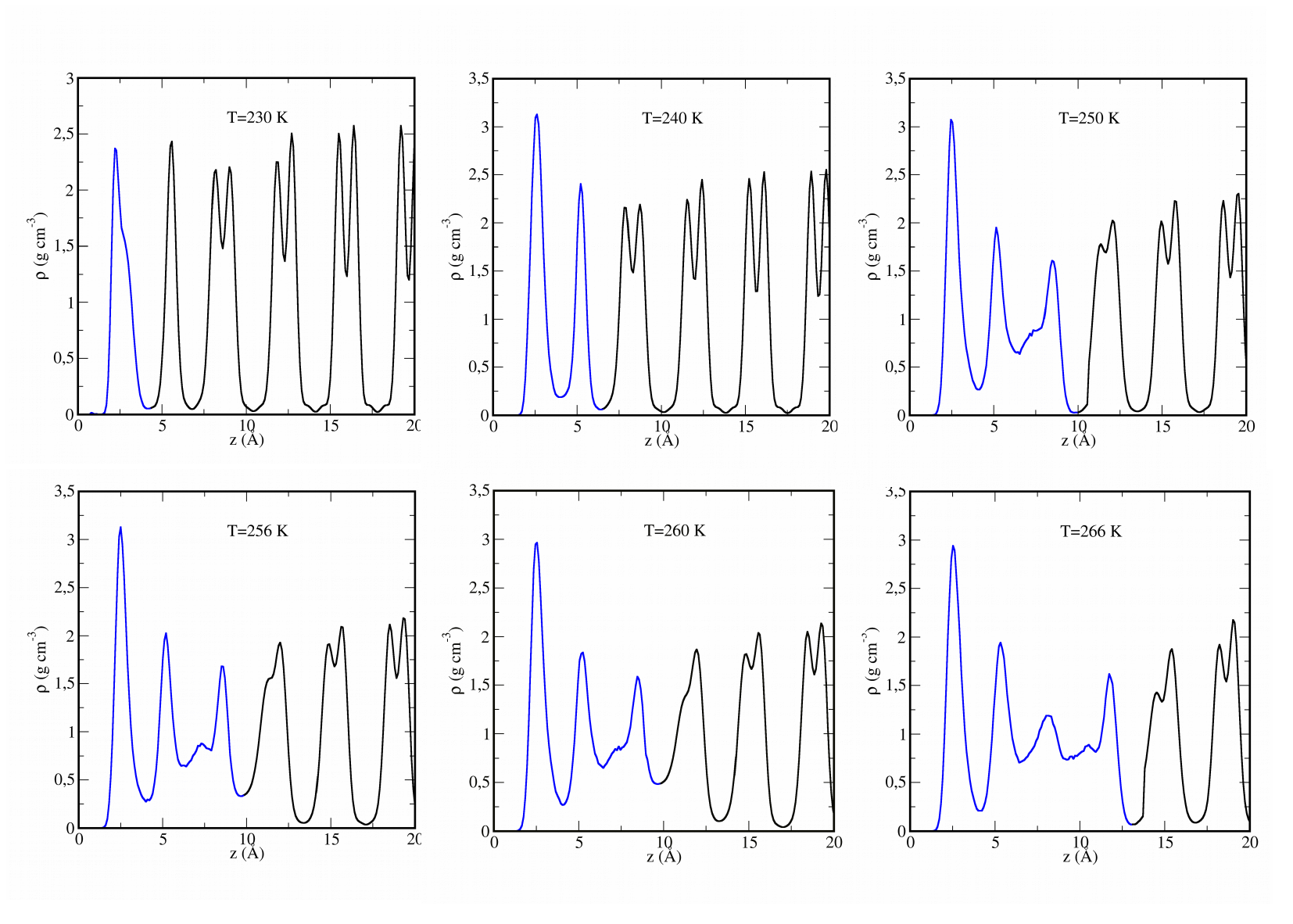} 
\end{center}
\caption{\label{fig:density_p1_f4}
   Density profiles along the  $p=1$~atm isobar  for the hydrophilic system with $\theta=50^\circ / f=4$. The continuous line is divided in two,
   with blue color depicting the region where liquid-like water is the majority phase and black color where ice is the majority phase. 
}
\end{figure}

\begin{figure}
\begin{center}
   \includegraphics[width=0.85\textwidth]{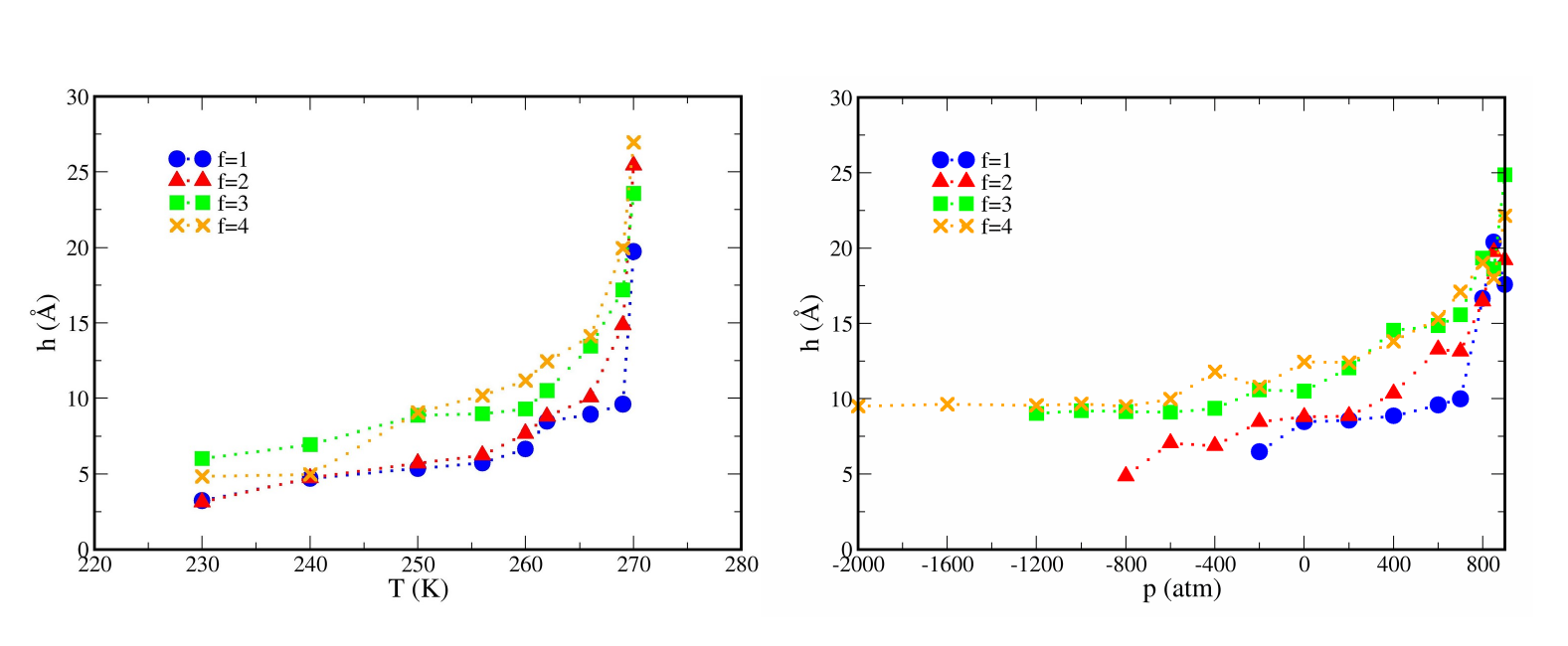} 
\end{center}
\caption{\label{fig:h_v_T_and_p}
  {\color{black}  Film thickness $h$ as a function of temperature (a) and pressure (b).}
}
\end{figure}

Figure~\ref{fig:density_p1_f1} shows a number of density profiles obtained along the $p=1$~atm isobar for a wall strength $f=1$ corresponding to a highly hydrophobic wall. The profiles display the mass density of H$_2$O molecules as a function of the distance $z$ away from the average location of the substrate. The solid phase is easily recognizable from the sharp doublet peaks, which correspond  to stacks of chair hexagons typical of ice Ih oriented perpendicular to its basal plane (also known as bilayers). The signature of premelting can be recognized as a  loss of the double peak structure of the density profile in the vicinity of the wall.\cite{baran22} We have checked this by calculating explicitly the density of liquid-like and solid-like water molecules, and confirming that indeed, the lack of double peaks is mostly attributed to the density of liquid-like molecules. To show this concisely and to avoid overcrowded figures, the part of the full 

Along with the density profiles for $f=1$, corresponding to a substrate with contact angle of ca. 120$^\circ$, Figure~\ref{fig:density_p1_f1} also displays density profiles for ice in contact with its vapor. This can be considered as the limit of a hydrophobic substrate, as the 'substrate' here is merely the vapor phase, and corresponds to an effective hydrophilicity parameter of $f=0$. The figure is striking, as it shows that  the interface of ice with a substrate with contact angle of 120$^\circ$ is very similar  to that of the ice-vapor interface. Therefore, the ice-vapor interface, which has been much studied, can serve as a proxy for the structure of interfacially premelted films on highly hydrophobic substrates. In the inssuing discussion, however, we will show that this is not quite the case when the substrates gradually become more hydrophilic.

Indeed, increasing the wall strength changes the situation significantly, both as regards the extent of premelting, and the height of the adsorption peak next to the wall. This can be seen in Figure~\ref{fig:density_p1_f4}, where we show density profiles for wall strength $f=4$, corresponding to a partially hydrophilic wall with contact angle of about $\theta=50^{\circ}$. At the lowest temperature, $T=230$~K, only one adsorption peak of liquid water is present, similar to wall strength $f=1$, but the liquid molecules are strongly adsorbed and the peak is now as high as the bilayer peaks of the solid phase. As the temperature increases, the premelting layer grows in extent  up to about four full molten bilayers, while the wall adsorbed peak increases in height significantly above the value found for bilayer peaks. 

The results appear to be consistent with experiments performed on widely different solid substrates, which show evidence of increased quasi-liquid layer thickness as the temperature raises for a wide range of materials such as glass, clays, talc, graphite and silicon oxide.\cite{maruyama92,gay92,furukawa93,ishizaki96,pittenger01,engemann04,schoder09,li19} According to our results, the thickness is small at T=230~K or below, and reaches nanometer thickness a few kelvin away from the melting point, in rough agreement with several of the experimental observations.\cite{pittenger01,engemann04,schoder09,li19} This tendency is confirmed by the study of the film thickness as a function of temperature, displayed in Fig.~\ref{fig:h_v_T_and_p}-a, which shows how the thickness evolves from somewhat below 1~nm to more than 2~nm close to the melting point.\cite{sun22,zhu23}

Our observation that the film thickness increases as the system becomes hydrophilic is in agreement with findings by Pittenger et al.,\cite{pittenger01} and with previous computer simulation studies,\cite{sun22,cui24} but at odds with conclusions from Li et al.\cite{li19} and Emelyanenko et al.\cite{emelyanenko20} This need not be a serious discrepancy, however, because experiments test the effect of hydrophilicity by studying widely different substrates, while our study is performed over very simplified model substrates whose hydrophilicity can be tuned in a continuous fashion. Furthermore, the measure of hydrophilicity based on contact angles usually has considerable uncertainties.\cite{santiso13}

\begin{figure}
\begin{center}
   \includegraphics[width=0.85\textwidth]{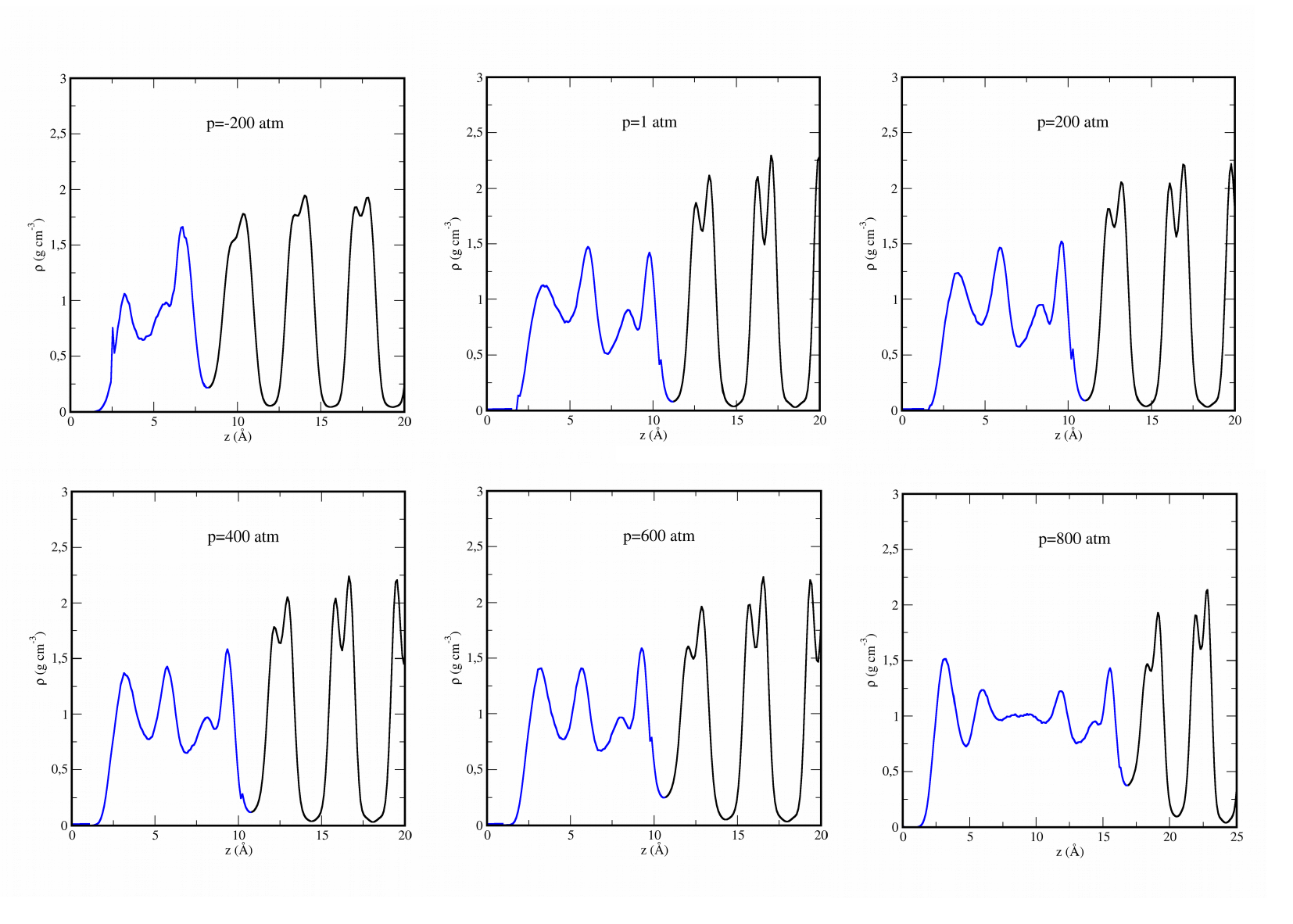} 
\end{center}
\caption{\label{fig:density_T262_f1}
   Density profiles along the $T=262$~K isotherm for the hydrophobic system with $\theta=120^\circ / f=1$. The continuous line is divided in two, 
   with blue color depicting the region where liquid-like water is the majority phase and black color where ice is the majority phase. 
}
\end{figure}

\begin{figure}
\begin{center}
   \includegraphics[width=0.85\textwidth]{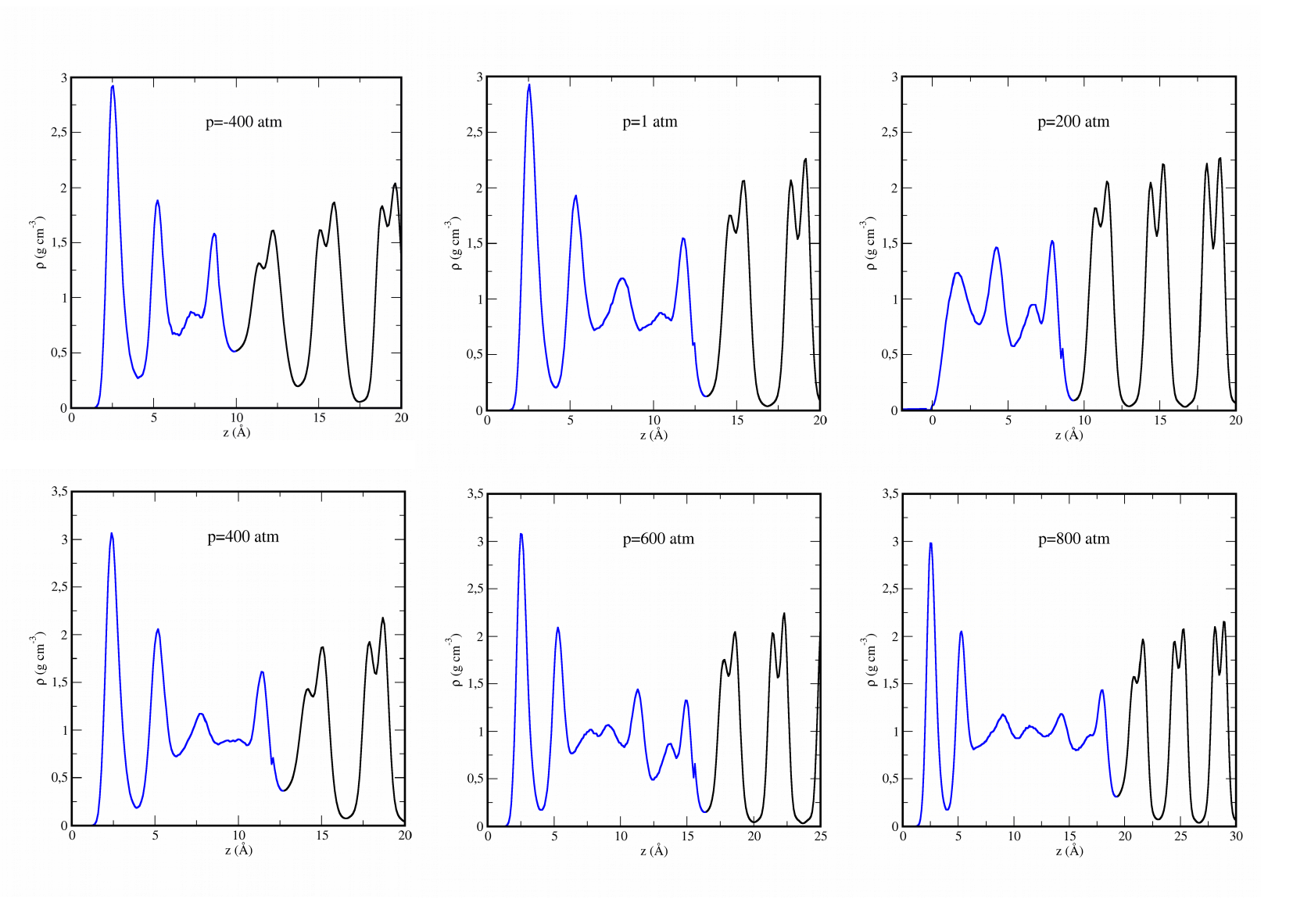} 
\end{center}
\caption{\label{fig:density_T262_f4}
  Density profiles along the $T=262$~K isotherm  for the hydrophilic system with $\theta=50^\circ / f=4$. The continuous line is divided in two,
   with blue color depicting the region where liquid-like water is the majority phase and black color where ice is the majority phase. 
}
\end{figure}

The increase of liquid adsorption is obviously related to the approach of ice to its melting point. But a situation somewhat less appreciated is that, by the same token, the extent of premelting should increase not only by increasing the temperature along an isobar, but also, by increasing the pressure along an isotherm.

Indeed, due to the anomalous behavior of water, the melting line exhibits a negative slope, so that pressurizing ice drives it closer to the liquid state (c.f. Fig.\ref{fig:coexistence}). This interesting possibility can be accomplished in a number of
practical situations, such as ice skating, but unfortunately, has not been studied experimentally to the best of our knowledge.  

To shed some light on the problem, we test the pressure effect in Figures~\ref{fig:density_T262_f1}-\ref{fig:density_T262_f4}, which exhibit density profiles at different pressures along the T=262~K isotherm for wall strengths $f=1$ and $f=4$. 

The results show how the density profiles evolve with increasing pressure in a way that very much resembles the evolution of density profiles with increasing temperature. 
Indeed, as the pressure increases we see that ice bilayers gradually melt, increasing the thickness of the quasi-liquid layer. Upon reaching 800~atm, which is close to our estimated melting pressure for the TIP4P/Ice model, the premelting layer appears to extend about five molten bilayers for the hydrophobic substrate ($f=1$), and up to six for the hydrophilic substrate ($f=4$). This tendency can be confirmed at a coarser scale from plots of the premelting film thickness as a function of pressure (Figure~\ref{fig:h_v_T_and_p}-b).

\subsection{4.2.  Equivalence between states on an isotherm with states on an isobar}

\label{sec:mapping}

\begin{figure}
	\begin{center}
		\includegraphics[width=0.65\textwidth]{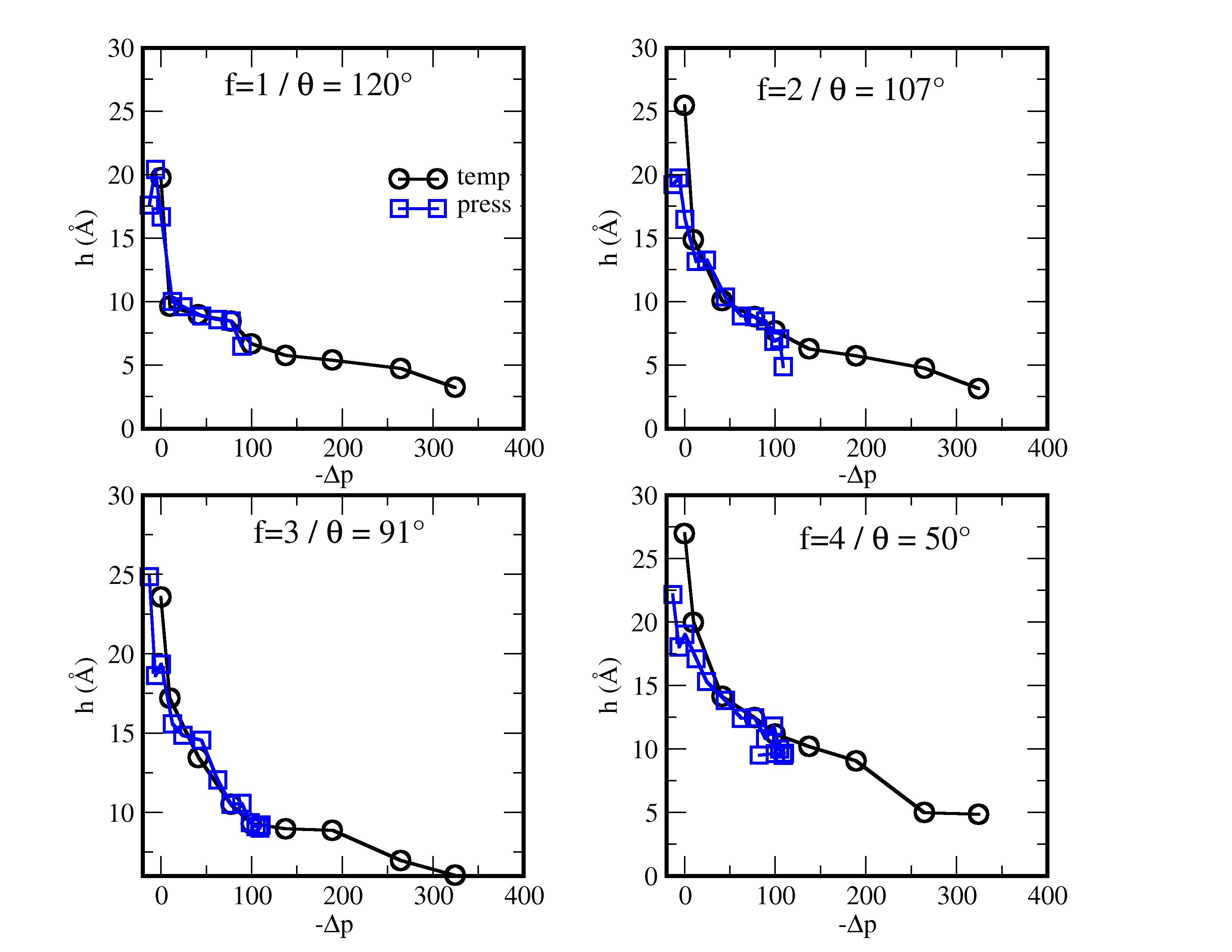} 
	\end{center}
	\caption{\label{fig:h_v_Pi_all_f} Premelting thickness for state at diffferent temperature and pressure plotted as a function of the single variable $\Delta p=p_w-p_i$. Note how the results from Fig.~\ref{fig:h_v_T_and_p} approximately collapse into one single curve. 
	}
\end{figure}

From the results performed over both an isobar (as usually in experiment) or an isotherm (as performed here) it appears that the extent of premelting can be described somehow as a function of the distance from  the melting point, i.e. as a function of $(T-T_m)$ for results along an isobar, or as $(p-p_m)$ for results along an isotherm. But one can clearly notice by comparison of Figures~\ref{fig:density_p1_f1}-\ref{fig:density_p1_f4} and Figures~\ref{fig:density_T262_f1}-\ref{fig:density_T262_f4}, that there are density profiles along an isotherm which very much resemble those found along an isobar. For example, the profiles  at $T=266$~K, $p=1$~atm (Figure~\ref*{fig:density_p1_f4})  look rather similar to those found at $T=262$~K, $p=400$~atm (Figure~\ref*{fig:density_T262_f4}). Similarly, one can identify by comparison of  panels a) and b) in Figure~\ref{fig:h_v_T_and_p}  thermodynamic states along the isobar which exhibit similar film thicknesses as those found along the isotherm.

An interesting question is, given that experiments along an isobar appear to be more readily achievable than those along an isotherm, is it possible to map ones on to the other? i.e., is there some way to map the states obtained along an isobar to the states obtained along an isotherm? And in that case, how does one map the values of $(p-p_m)$ of an isotherm with corresponding values of $(T-T_m)$ along the isotherm?

The answer to this question lies on the interface potential concept discussed in Section~2.1.\ref{sec:theory}. 
Indeed, there we saw that, under the reasonable assumption that $g(h)$ is only weakly state dependent, one expects, according to Eq.(\ref{eq:eqcond}), that states with equal value of $\Delta p=p_{water}-p_{ice}$ will exhibit similar quasi-liquid layer structure and thickness.

As a first check, Figure~\ref{fig:h_v_Pi_all_f} displays all film thicknesses obtained in this work, either along an isotherm or along an isobar, as a function of the single variable $-\Delta p$. One can clearly see here that when plot in this way, all the results fall in a similar scale. Moreover, it is clear that, except for the results with $\Delta p$ close to zero, the results collapse onto one single master curve. This shows that, indeed, to a semi-quantitative extent, it is possible to map properties of systems $(p-p_m)$ away from the melting point along an isotherm with points that are $(T-T_m)$ away from the melting point along an isobar. This allows one to predict film thicknesses $h(T,p_{ice})$ for arbitrary values of the imposed temperature and ice pressure in terms of simulation (or experimental) results obtained at other conditions.

\begin{figure}
\begin{center}
   \includegraphics[width=0.65\textwidth]{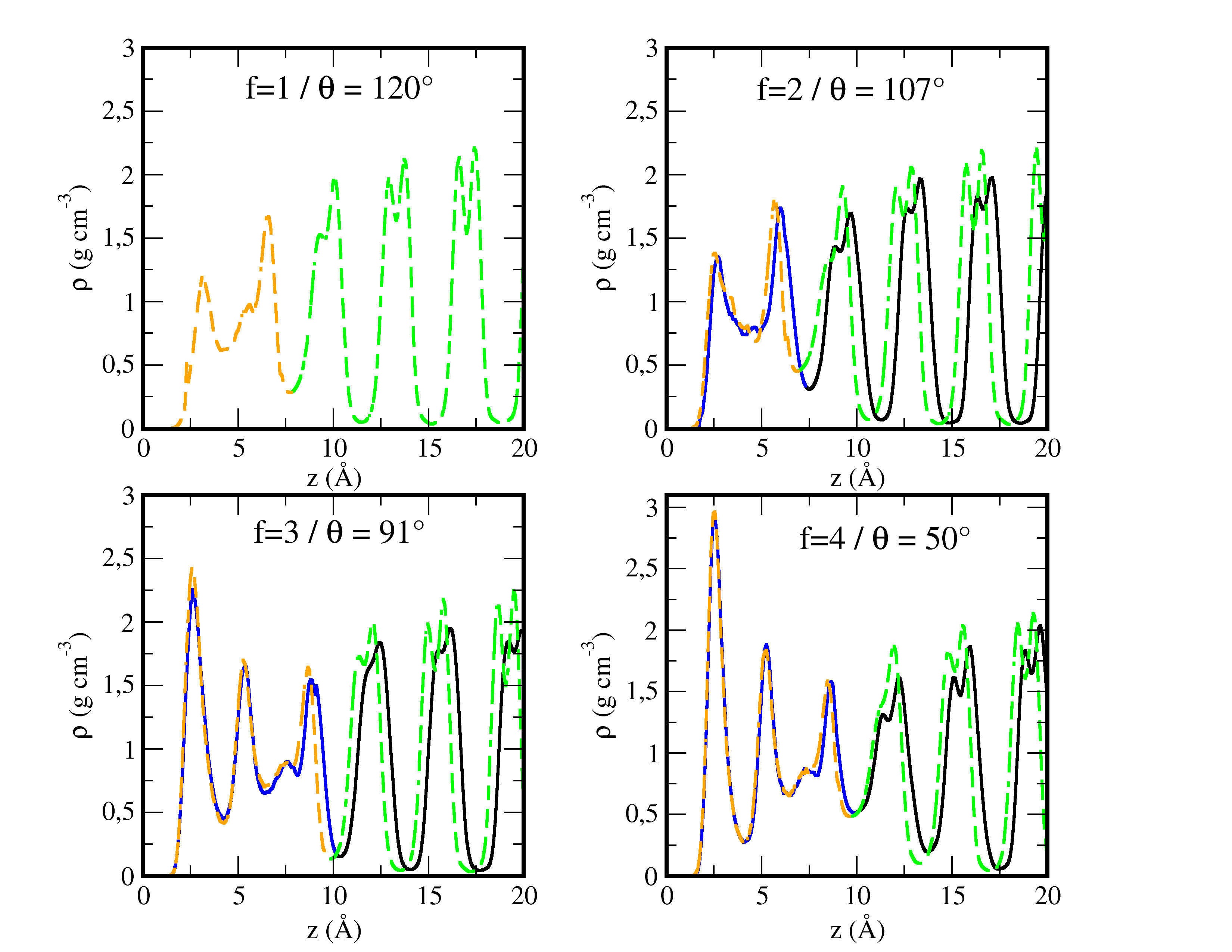} 
\end{center}
\caption{\label{fig:T262@p_400_vs_T260@p1}
   Comparison of density profiles obtained at  $T=262$~K and $p=-400$~atm (solid lines), with those obtained at $T=260$~K and $p=1$~atm (dashed lines). 
   The density profiles are divided in two,
   with blue and orange colors depicting the region where liquid-like water is the majority phase and black and green colors where ice is the majority phase.  
}
\end{figure}

\begin{figure}
\begin{center}
   \includegraphics[width=0.65\textwidth]{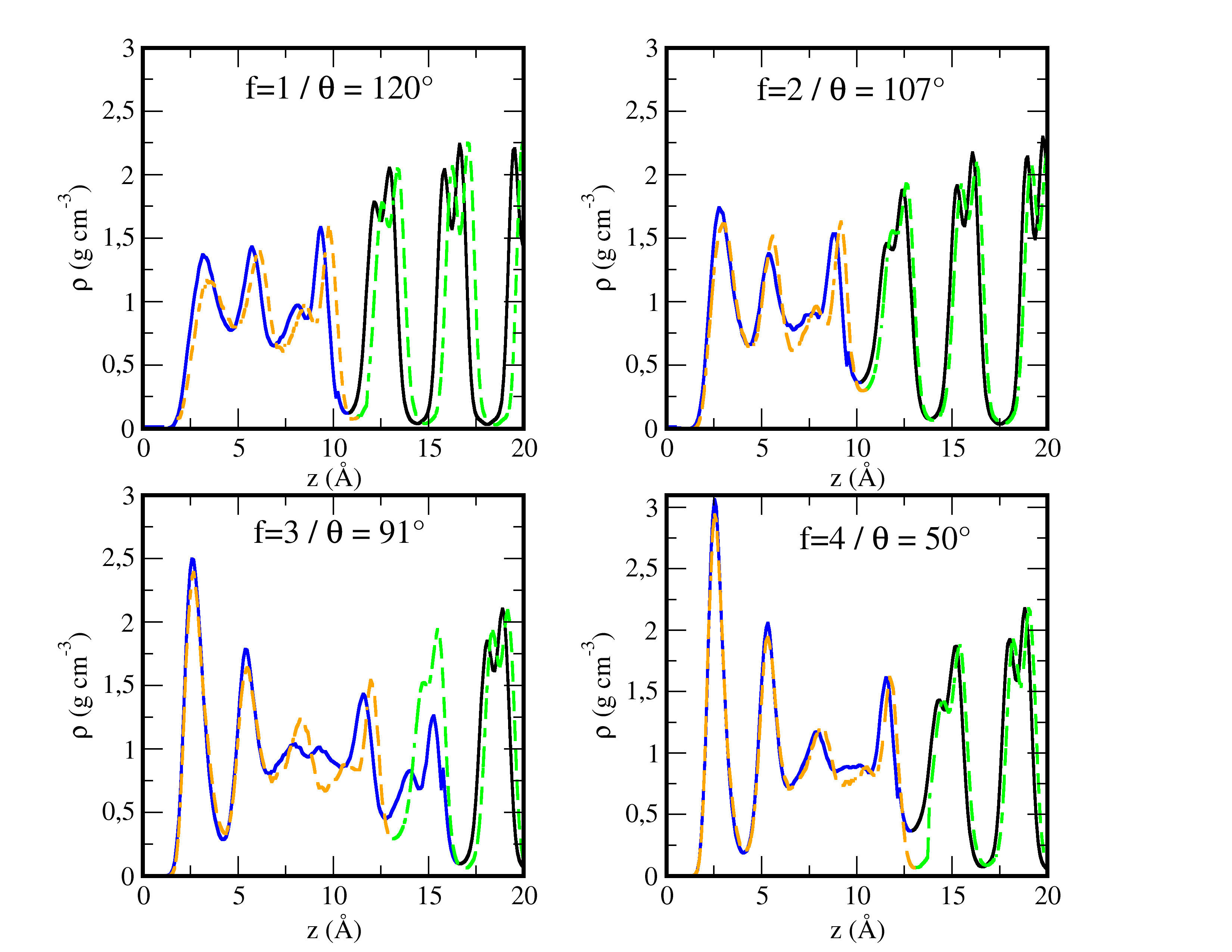} 
\end{center}
\caption{\label{fig:T262@p400_vs_T266@p1}
   Comparison of density profiles obtained  at $T=262$~K and $p=400$~atm (solid lines) 
   with those obtained at $T=266$~K and $p=1$~atm (dashed lines). 
   The density profiles are divided in two regions,
   with blue and orange colors depicting the region where liquid-like water is the majority phase and black and green colors where ice is the majority phase.  
}
\end{figure}

\begin{figure}
\begin{center}
   \includegraphics[width=0.65\textwidth]{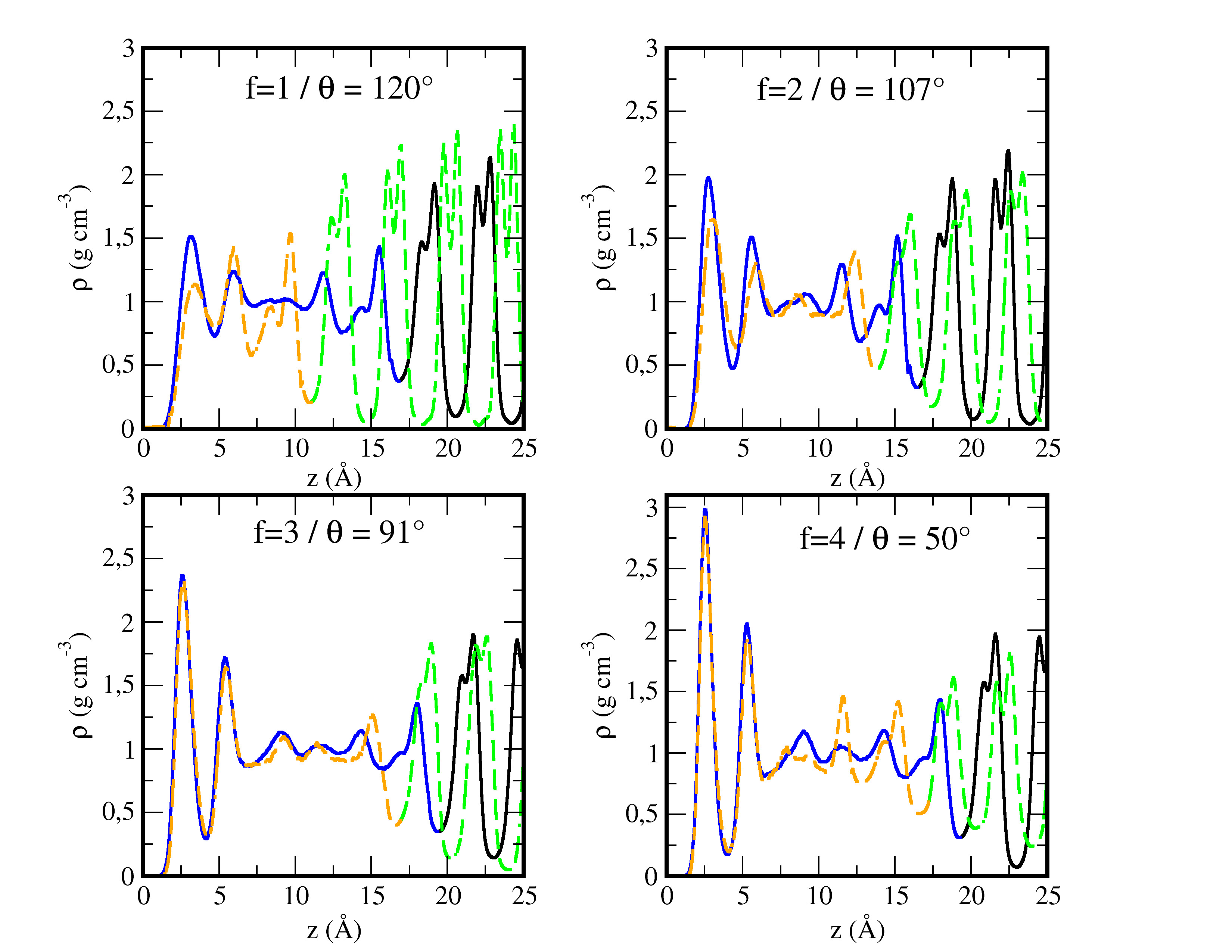} 
\end{center}
\caption{\label{fig:T262@p800_vs_T269@p1}
   Comparison of density profiles obtained 
   at $T=262$~K and $p=800$~atm (solid lines) 
   with those obtained at $T=269$~K and $p=1$~atm (dashed lines). 
   The density profiles are divided in two,
   with blue and orange colors depicting the region where liquid-like water is the majority phase and black and green colors where ice is the majority phase. 
}
\end{figure}

Of course, the film thickness is an important property for the characterization of a quasi-liquid layer, but one could conceive two different states with similar film thickness but very different interface structure. 

Our results appear to suggest that, for the rather large range of conditions studied here (between -2000 to 900~atm and 230 to 269~K), it is not this situation, but rather, the contrary which holds rather accurately. 
This is illustrated in Figures~~\ref{fig:T262@p_400_vs_T260@p1}
\ref{fig:T262@p400_vs_T266@p1}~\ref{fig:T262@p800_vs_T269@p1}. 

To show this, consider first our simulations at ($T=266$~K, $p=1$~atm), and ($T=262$~K, $p=400$~atm), which are found at rather different thermodynamic conditions, but have very similar values of $\Delta p(T,p_{ice})$ (c.f. $\Delta p=41.49$ vs $\Delta p=45.04$, Supplementary Table~5,6). A comparison of the full density profiles of these two systems is shown in Figure~\ref{fig:T262@p400_vs_T266@p1}. The results for profiles differing in 400~atm are indeed in remarkable good agreement for the four wall strengths studied ($f=1$ to $f=4$).

Now, consider the comparison between density profiles obtained at ($T=260~$K, $p=1$~atm) and ($T=262~$K, $p=-400$~atm), which despite the large pressure difference and the metastability of the latter state, share also similar values of $\Delta p(T,p_{ice})$ (c.f. $\Delta p=99.86$ v $\Delta p=98.48$, Supplementary Table~S5,S6)). The results of Figure~\ref{fig:T262@p_400_vs_T260@p1}  show how the density profiles of these states are again remarkably similar, except for the hydrophobic wall $f=1$, which is unstable at $-400$~atm and detaches from the wall due to the tension. 

The conclusion from Figures~\ref{fig:T262@p_400_vs_T260@p1}~\ref{fig:T262@p400_vs_T266@p1} is that the results performed along an isobar at the convenient pressure of $p=1$~atm can serve to characterize not only the film height, but the full structure of the quasi-liquid layers in a range of pressures from -400 to 400~atm and arbitrary temperatures.

Given these findings, one  wonders whether the properties of premelting films at the melting line, which always have $\Delta p=0$ by virtue of the coexistence condition, should exhibit the same structure along the full melting line. We can test this here for our results ($T=269$~K, $p=1$~atm), which are less than $1~K$ away from the melting point, with those obtained for the state $(T=262$~K,$p=800$~atm), which lies also slightly below melting point of the TIP4P/Ice model according to our calculations. The comparison of Figure~\ref{fig:T262@p800_vs_T269@p1} shows that the premelting films at these two different state points differ to a larger extent from those compared in Figures~\ref{fig:T262@p_400_vs_T260@p1}~\ref{fig:T262@p400_vs_T266@p1}, particularly for the hydrophobic wall ($f=1$), but are still at least qualitatively similar.

We conjecture that the properties of the quasi-liquid layer remain very similar along the melting line, but will eventually differ when the the states that are compared differn in pressure by a large extent. In this case, it seems that a difference in pressure of 800~atm between the choice of states along the melting line is close to the limit where this comparison can be made meaningfully.

However, the significance of the mapping implied in Figure~\ref*{fig:h_v_Pi_all_f} should not be understated, as it allows to provide quantitative or at least qualitative results for the quasi-liquid layer thickness over a wide range of thermodynamic conditions of relevance to ice friction, adhesion and regelation.

\subsection{4.3. Interface potential and Disjoining pressure curves}

 In this section, we show how the combination of the surface thermodynamics embodied in the concept of interface potential together with knowledge on surface intermolecular forces provides a powerful framework for the characterization of the premelting behavior, including the mapping of liquid film structure between different thermodynamic states discussed in the previous section.

In the absence of dissolved ions, the interface potential can be described in terms of two additive contributions.\cite{derjaguin87,churaev88,evans94,henderson05,israelachvili11} One is the short range, or structural contribution, which decays exponentially fast. It results  from packing correlations and other short range  interactions such as hydrogen bonds. The other is the long range or van der Waals contribution, which results from long range dispersion forces and decays algebraically slow.\cite{parsegian05,israelachvili11} 

A minimal model to account for such interactions is:
\begin{equation}\label{eq:expandvdw}
g(h) =  C e^{-\kappa h} - \frac{A}{12\pi h^2}
\end{equation}
where $C$ and $\kappa$ are constants which determine the surface free energy and length scale of the short range contribution, while, $A$-the Hamaker constant-sets the scale of the van der Waals interactions.

We emphasize this is a {\em minimal} model. Both short and long range contributions are necessary to provide for the correct behavior of the interface potential. This contrasts with many experimental and theoretical work,\cite{maruyama92,gay92,furukawa93,ishizaki96,pittenger01,wettlaufer06,bostrom17,bostrom19,esteso20,style23,engemann04,schoder09,li19} were either the short range,\cite{engemann04,schoder09,li19} or the long range terms alone, \cite{maruyama92,gay92,furukawa93,ishizaki96,pittenger01,wettlaufer06,bostrom17,bostrom19,esteso20,style23} are employed to describe the system's behavior.

In particular, the distinction between exponential and algebraic decay is important. For large film thicknesses, the short range contribution has vanished completely, and the long range contribution fully dictates the behavior of the interface potential. However, as we shall soon see, in  the very relevant range of units of nanometer, the short range forces {\em dominate} completely the system's behavior.

These expectations can  can be tested experimentally by measuring film thickness as a function of temperature. Indeed, using the model of Eq.\ref{eq:expandvdw} together with  Eq.\ref{eq:eqcond} and Eq.\ref{eq:dpisobar}, it is possible to solve for $h=h(T-T_m)$. In the event that $h$ is very large, the algebraic term dominates. Then, provided $A$ is negative,  one finds that the film thickness should diverge as a power law, $h\propto (T_m-T)^{1/3}$. Alternatively, if the thickness is in the range of one nanometer or less, and the van der Waals forces can be neglected, the film thickness will exhibit a logarithmic divergence,  $h\propto -\ln(T_m-T)$ provided $C$ is positive. However, it must be emphasized that in the range of thicknesses where both contributions are important simultaneously, it is not possible to represent the film growth in terms of either the purely logarithmic or purely algebraic models. 

Finally, as a word of caution we note that  we expect this model to work for films up to ca. 100~nm thick. For even thicker films, things become tricky, because the electronic dipole fluctuations  embodied in the Lennard-Jones potential are suppressed due to quantum-electrodynamic effects.\cite{dzyaloshinskii61,parsegian05,macdowell17}  As a result, the Hamaker 'constant' becomes $h$ dependent and  can even exhibit a sign reversal.\cite{wilen95,macdowell17,esteso20} However, this occurs at such large thicknesses that it corresponds to equilibrium films that can only be probed at temperatures of a milli-Kelvin away from the melting point, so we may safely ignore this complication under most practical circumstances.

\subsubsection{Long range forces prevent complete surface melting}

Let us first consider the long range van der Waals contribution, which dominates the behavior of the premelting film for large film thicknesses. Particularly, whether the film thickness has a propensity to increase unbound and attain infinite thickness (i.e. to wet or not to wet the substrate completely) is dictated by the sign of the Hamaker constant. 

Fortunately, the Hamaker constant is well understood and can be predicted accurately by theoretical considerations alone. For this purpose, one can resort to the rather sophisticated 
quantum-electrodynamic theory of Lifshitz,\cite{dzyaloshinskii61,parsegian05} which has been often applied to understand ice premelting in a number of different contexts. \cite{elbaum91b,wilen95,bostrom17,bostrom19,esteso20,luengo20} For our purpose, however, it suffices to consider a much simpler theoretical framework borrowed from a model of additive dispersive interactions.\cite{hamaker37,dietrich91,macdowell17,evans19} Applying such results for the special case of ice premelting, we find:
\begin{equation}\label{eq:hamaker}
A = {4\pi^2 \varepsilon\sigma^6\left(\rho_{S}\sqrt{f} - \rho_{w}  \right )
 \left(\rho_{i} - \rho_{w}  \right )
}
\end{equation}
where $\epsilon$ and $\sigma$ are the Lennard-Jones parameters of the water model,
 $\rho_{S}$, $\rho_{i}$ and $\rho_{w}$ are the number densities of the model wall, ice and water, respectively and $f$ is the parameter dictating the 
hydrophilicity of the wall.\footnote{Again notice this approximation works for van der Waals interactions related to high frequency electronic dipole fluctuations, corresponding to the algebraic decay in the scale of 100~nm or less. Such effects are modeled effectively in classical computer simulations by the Lennard-Jones potential. At larger thicknesses we note that the van der Waals interactions become dominated by low frequency dipole fluctuations, which are not grasped by the simple expression above and need a more elaborated quantum-electrodynamic theory.\cite{parsegian05,israelachvili11}}

Taking into account the properties of our model system at phase coexistence we find  Hamaker constants that  fall in the range of $10^{-21}$~J (i.e. one zJ) as is usual (c.f. Table~\ref{tab:hamaker-fit}). More significantly, they all have positive sign. This means that at large values of $h$, the interface potential will be negative. The way to interpret this is that a film of limited thickness will have lower free energy, and be more favorable, than one of large thickness. Therefore, in our systems the van der Waals forces will not allow for the {\em complete} surface melting of ice, which must thus exhibit quasi-liquid layers of limited thickness up to milli-Kelvins away from the melting point (a sign reversal of the Hamaker constant due to low frequency dipole fluctuations could change this scenario, but this would occur for very thick films beyond the hundreds of nanometers that can only be attained extremely close to the melting point). Furthermore, we find that contrary to some expectations,\cite{emelyanenko20} the van der Waals forces acting against the growth of premelting layers are stronger for the hydrophobic materials than for the hydrophilic ones. Indeed, inspection of Table ~\ref{tab:hamaker-fit} shows that the Hamaker constant for the most hydrophobic material is more than one order of magnitude larger than that of the most hydrophilic material. This behavior actually can be read off directly from Eq.\ref{eq:hamaker}, which shows that, the smaller is $f$, the larger is the absolute value of  the first parenthesis in the right hand side of the equation, and hence, the more positive is the  Hamaker constant.

\begin{table}[h!]
    \centering
    \begin{tabular}{c|c|c|c|c|c}
        f & $\theta$ & C (mJ/m$^2$) & $\kappa$ (\AA$^{-1}$)  & $A$~(zJ) & $h_e$ (\AA)\\
        \hline
        0 (basal) &  -  & 169   & 0.738 & 5.07 &  12 \\
        0 (pI)  & -     & 170   & 0.880 & 5.07 &  9.7 \\
        1       & 120 & 31.330 & 0.323   & 2.27  & 31  \\
        2       & 107 & 30.322 & 0.258   & 1.36  &  44 \\
        3       & 91  & 66.079 & 0.289   & 0.66  &   45 \\
        4       & 50  & 40.969 & 0.166   & 0.071 &  100 \\
        > 4.1   &  < 50 & -    & -        & < 0   & $\infty$ \\
    \end{tabular}
    \caption{Fitting parameters to the short-ranged contribution of the disjoinig pressure (orange curves in Fig.~\ref{fig:disjoining-contributions-temp}). 
    The values of the Hamaker constant are given
    for temperature $T=270$~K at $p=1$~atm. Also included are the equilibrium premelting film thickness at bulk coexistence as determined from the minimum of the model interface potential.}
    \label{tab:hamaker-fit}
\end{table}

The Table~\ref{tab:hamaker-fit} discusses the value of Hamaker constants as evaluated at the melting point.  However, at odds with some simpler systems, the presence of the second parenthesis in Eq.~\ref{eq:hamaker}, i.e., the density difference between water and ice, implies that the Hamaker constant changes significantly upon changing the thermodynamic state e.g., for $f=4$, and states along the $p=1$~atm isobar, $A$ changes from $0.0715$~zJ at $T_m$ to $-0.0462$~zJ at $T=230$~K. Similarly, for states along the $T=262$~K isotherm, the Hamaker constant changes from $0.0410$ at $p=1$~atm to $0.3089$ at $p=800$~atm for the hydrophilic system, but can eventually vanish and reach negative values for ice under tension 
(cf. Tables S3, S4 in Supplementary Material). 

\subsubsection{Short range forces favor premelting}

Our results emphasize how important it is to use the minimal model of Eq.(\ref{eq:expandvdw}) to properly describe the premelting behavior. How can the system's propensity to exhibit significant premelting that we have discussed in the previous sections be accommodated in a description where the van der Waals tail prevents the film from growing\dots?

The reason is that there also are short range forces, which, in this case, must somehow promote growth of a premelting layer at short range. 

Unfortunately, our understanding of the short range behavior is more limited than that of the van der Waals contribution. From theoretical considerations, it is well understood that the parameter $\kappa$, which dictates the range of the exponential decay is constrained to fall in the scale of the molecular diameter.\cite{chernov88,evans94,henderson05} However,  it is more difficult to constrain $C$,   which typically falls in the scale of mJ/m$^2$, but can be either positive or negative.

In absence of additional theoretical guidance, simulation or experimental results provide insight onto  the related {\em disjoining pressure}, which is defined as  minus the derivative of the interface potential (c.f. Eq.~\ref{eq:eqcond} and accompanying text).

\begin{figure}
\begin{center}
   \includegraphics[width=0.65\textwidth]{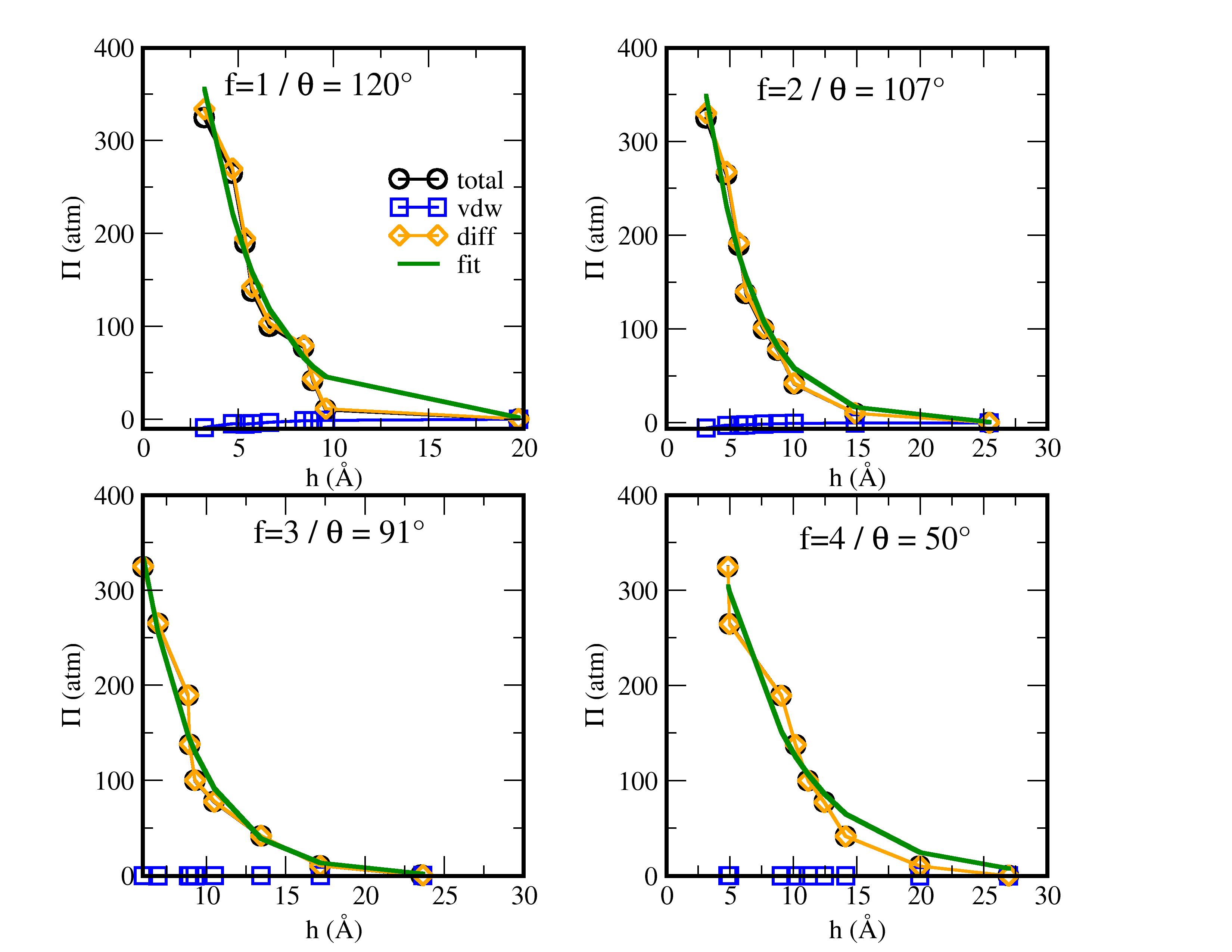} 
\end{center}
\caption{\label{fig:disjoining-contributions-temp}
   Disjoinig pressure as a function of premelting film thickness obtained along an isobar $p=1$~atm. Full disjoining pressure is shown as black circles,
   whereas the vdW forces are shown in blue squares and short-ranged contributions are displayed in orange, respectively. An exponential fit to the short-ranged contribution is presented as a green solid line. 
}
\end{figure}

Figure~\ref{fig:disjoining-contributions-temp} displays  the disjoining pressure of interfacially premelted films as obtained from results of our simulations along the isobar.  The results show positive disjoining pressures with negative slope, implying the favorable formation of premelting films of finite extent within the range of conditions studied. This in turn implies that the constant $C$ must be positive. 

Together with the full disjoining pressure curves obtained from simulation, Figure~\ref{fig:disjoining-contributions-temp} displays the van der Waals contributions estimated from Eq.~\ref{eq:hamaker} (c.f. Supplementary Material for the calculations).  The results show that in the range of premelting film thicknesses studied here, the van der Waals contributions are  a very small fraction of the full disjoining pressures in all cases, not exceeding more than ca. 5~\% in the range of film thicknesses studied.   

In order to show this more clearly, Figure~\ref{fig:disjoining-contributions-temp} displays in orange color the difference between the disjoining pressure obtained from simulation, with the theoretical predictions for the van der Waals forces. The resulting curve corresponds to the purely short range contribution, and differs only to a negligible extent from the full disjoining pressure. Therefore, we find that, against common believe,\cite{maruyama92,gay92,furukawa93,ishizaki96,pittenger01,garcia08,meyer24}  van der Waals forces in the the range of units of nanometer thickness are only a very small fraction of the total surface intermolecular forces.

The significance of the short range contribution explains why previous attempts to fit the experimental results of film thickness with a model of long range van der Waals forces did not provide the expected power law behavior discussed in the previous section.\cite{maruyama92,gay92,furukawa93,ishizaki96,pittenger01} In the light of our results, we see that the  problem must have resulted from the attempts to fit the logarithmic behavior that results from the dominant short range forces, with a power law corresponding to long range forces. In fact, the few studies which instead resorted to a model of short range decay showed far better agreement with the expected logarithmic law.\cite{engemann04,schoder09,li19} 

Of course, for very large thicknesses the exponential term becomes completely suppressed, and then 100~\% of the disjoining pressure stems from the van der Waals forces only. However, such thicknesses can only be achieved for systems exceedingly close to coexistence, as the equilibrium condition, Eq.(\ref{eq:eqcond}), requires $\Delta p\approx0$. By use of Eq.(\ref{eq:dpisobar}), this can only be achieved in close proximity to  the melting line, under a very fine temperature control which has usually not been attained in experiment. This circumstance was recognized in Ref.\cite{cahn92} 

Now, however small the van der Waals forces are, they are crucial in determining the location of the minimum of the interface potential. Thus, in order to assess the location of the minimum, we  fit the short range contribution to a single exponential function, as dictated by the model of  Eq.\ref{eq:expandvdw}. The fits, shown in Fig.~\ref{fig:disjoining-contributions-temp}, appear in reasonable agreement with the simulation results. Table~\ref{tab:hamaker-fit} collects all the parameters from the fit, together with the calculations for the  Hamaker constant. Included also for comparison are results obtained previously for the premelting of ice in contact with its vapor, which can be considered the limit of a hydrophobic wall with $f=0$.\cite{luengo22b,baran24b} This set of parameters  provides a minimal model of interface potentials for premelting films of differing hydrophilicity. 

The model can now be used to extrapolate the system's behavior beyond the range of thicknesses actually probed by simulation and used to determine the minimum of the interface potential. i.e., the equilibrium film thickness at exact bulk coexistence. The results obtained from the extrapolated interface model are shown in Table~\ref{tab:hamaker-fit} and exhibit an increase of the premelting film thickness as the substrate becomes more hydrophilic.
The equilibrium thicknesses that are obtained are already rather large, and imply our systems are close to wetting. Indeed, the minima of the interface potential are found for very small values that are in the range of $10^{-3}$~mJ/m$^2$, which is an extremely small fraction of the ice/water surfacetension. The shallowness of such minima can be readily seen by use of Eq.~\ref{eq:gmin2}, which, together with an ice-water surface tension of ca.~30~mJ$/$m$^2$, yield vanishing contact angles of less than 10$^{-3}$$^\circ$ for water droplets embedded between the substrate and ice. This is consistent with the actual premelting of ice in contact with its vapor, which we have already shown to be close to the complete wetting transition.\cite{llombart20,sibley21,luengo22b}

Actually, we can seek for the value of $f$ first leading to a vanishing Hamaker constant by equating  the first factor in the right hand side of Eq.~\ref{eq:hamaker} to zero. Using densities from our model, this shows that the Hamaker constant vanishes at $f=4.12$, which, from Table~\ref{tab:hamaker-fit} corresponds to substrates with a water contact angle of slightly less than 50$^\circ$. Therefore, our data implies that hydrophilic substrates with water contact angles less than 50$^\circ$ must all have negative Hamaker constants and therefore become completely wet at bulk ice-water coexistence.

\subsubsection{Mapping concept}

In section 4.2.\ref{sec:mapping} we have seen that the premelting film thickness and the film structure appear to be functions of the pressure difference $\Delta p$ alone. This observation has allowed us to predict the structure of one thermodynamic state from knowledge of the structure of different thermodynamic states with equal value of $\Delta p$. In wetting physics, this mapping is supported formally by the concept of {\em interface potential}, $g(h)$ discussed above. Indeed, the equilibrium condition, Eq.\ref{eq:eqcond} suggests that states with equal $\Delta p$ are expected to have similar premelting film thickness and similar properties, provided the interface potential is  {\em not} state dependent.

Whence, in order to understand to what extent does this concept hold, it is required to consider the
 dependence of intermolecular forces on the thermodynamic state of the system.

In the previous section, we have discussed the Hamaker constants, and considered its value in conditions close to the melting point. However, at odds with some simpler systems, the presence of the second parenthesis in Eq.\ref{eq:hamaker}, i.e., the density difference between water and ice, implies that the Hamaker constant changes significantly upon changing the thermodynamic state. e.g., for $f=4$, and states along the $p=1$~atm isobar, $A$ changes from $0.0715$~zJ at $T_m\approx 270$~K to $-0.0462$~zJ at $T=230$~K. Similarly, for states along the $T=262$~K isotherm, the Hamaker constant changes from $0.0410$ at $p=1$~atm to $0.3089$ at $p=800$~atm for the hydrophilic system, but can eventually vanish and reach negative values for ice under tension (c.f. Table S3 in Supplementary Materials at $p=-200$~atm and the pressures below). 

From the discussion above, we see that, in principle, $g(h)$ and the corresponding disjoining pressure, $\Pi(h)$, can be state dependent, as illustrated by the state dependence of the Hamaker constant. On the other hand, we have seen in Section~4.2.\ref{sec:mapping} that the mapping of premelting structure between states with different thermodynamics holds rather accurately, which can be expected to occur when the disjoining pressure is not significantly state dependent.

The solution to this conundrum lies once more in Fig.~\ref{fig:disjoining-contributions-temp}, which shows that the van der Waals forces contribute only to a small extent to the overall disjoining pressure. Accordingly, it must be concluded that the short range structural contributions are indeed very weakly dependent on the thermodynamic state. This makes sense, because such contribution stems mainly from harsh repulsive forces and strong hydrogen bonds, which are far less susceptible to changes in thermodynamic conditions. Such a circumstance is actually exploited to great advantage in all perturbation theories of the liquid state.\cite{mansoori69,barker67,weeks71,barker76}

\subsection{4.4. Adhesion tests}
\begin{figure}[ht!]
\begin{center}
\includegraphics[width=0.70\textwidth]{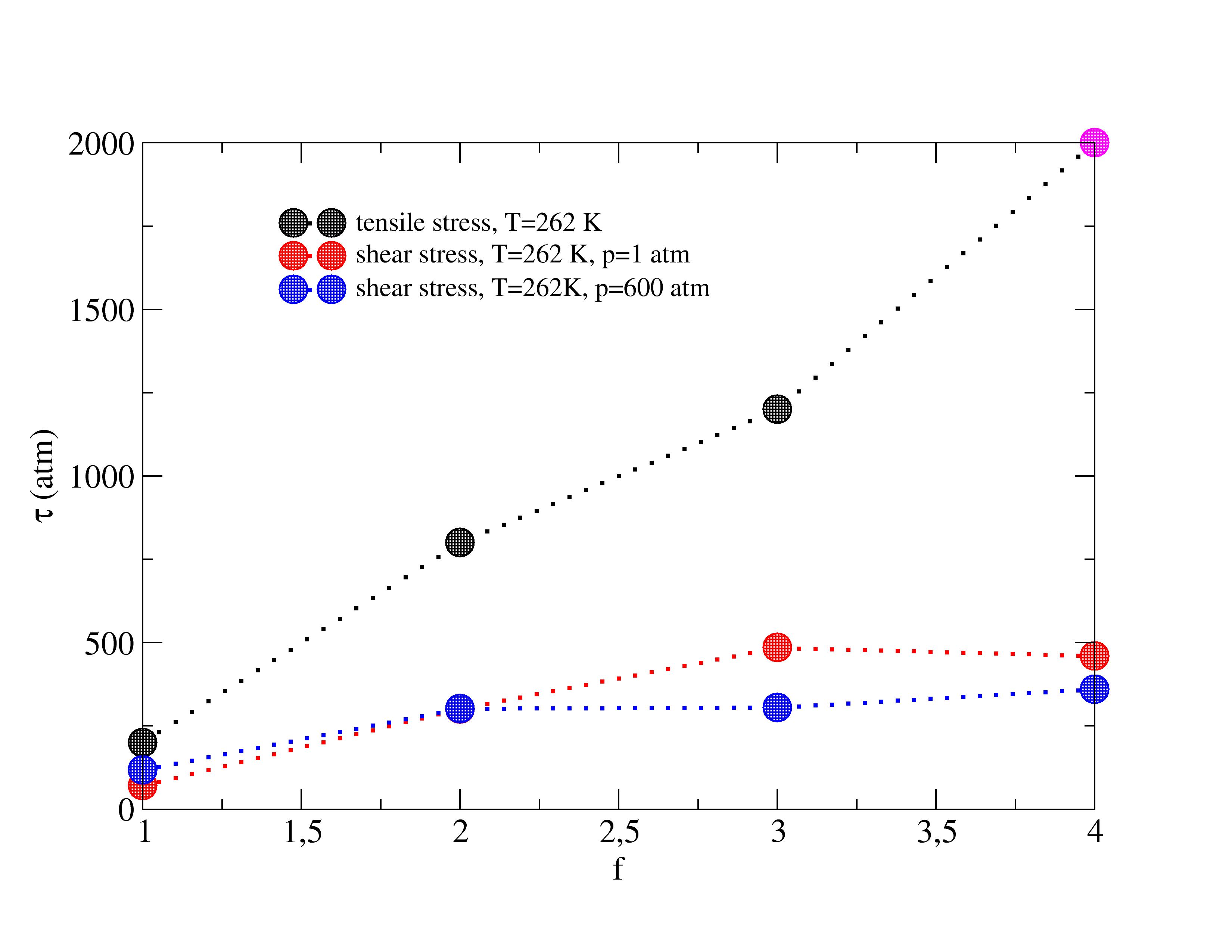} 
\end{center}
\caption{The relations of shear and tensile stress with respect to the 
wall strength $f$. 
The pink point in tensile stress curve for the system with $f=4$ at $p=2000$~atm emphasizes that the system remained stable for the whole duration of the simulation.  
\label{fig:shear-v-tensile}
}
\end{figure}

Whether in a weak or strongly premelted state, the quasi-liquid layer feels attractive interactions from the wall atoms which result in the adhesion of the ice sample to the wall.

This situation can be tested right away with the wall barostat in our simulations, as changing the direction of the force exerted by the wall atoms allows us to exert a pull rather than a compression on the ice sample. This  effectively corresponds to the simulation of a tensile adhesion test. Under such conditions, the system can remain in precarious equilibrium for some time. However, upon exercising sufficiently large negative pressure, ice detaches from the substrate, which corresponds effectively to an event of adhesive failure. Similar methodologies have been previously used to assess adhesion by computer simulations.\cite{xiao16,ronneberg20,sun22,cui24}

Indeed, several of the density profiles measured along the $T=262$~K isotherm (c.f. Section 4.1.\ref{sec:structure}) correspond to tests of ice under tension, which we reported there as a negative pressure. Our results show that such systems, which are actually in a state of precarious stability, could survive for the whole duration of the simulation, which amounts to 100~ns. However, the maximum tension that can be withstand by the wall depends very much on the wall strength $f$. 

The results of the tensile  adhesion strength tests are collected in Figure~\ref{fig:shear-v-tensile}. Tensile stresses  show a clear increase of adhesion with the hydrophilicity $f$, with values ranging from hundreds to thousands of atmospheres as the hydrophilicity increases. In fact, the tensile stress result for $f=4$ is a lower bound to the actual value, since the system remained stable for the whole duration of the simulation up to a pressure of $-2000$~atm. This results are in line with early measurements of ice adhesion, which show that adhesion by hydrophilic materials can be so strong that cohesive failure of the material occurs before the adhesive failure is ever achieved.\cite{raraty58,jellinek62} This behavior has been confirmed in many modern studies,\cite{liljeblad17,rehfeld24} and appears to occur also in computer simulations.\cite{sun22}

Together with the results for tensile strength tests, we include previous results of the shear stress measured while shearing an equilibrated premelting film at a sliding speed of $5$~m$/$s and two different pressures. Such results correspond effectively to shear adhesion strength tests of our sample, and appear to show a small influence of the perpendicular pressure on the shear stress.  

On qualitative grounds, our results  agree with adhesion tests, which show that tensile stress is significantly  larger than shear stress  under most circumstances.\cite{jellinek62,liljeblad17,rehfeld24} In line with the early measurements and modern studies, we also find adhesion to atomically smooth surfaces to be larger for hydrophilic than hydrophobic materials for either shear or tensile stress tests.\cite{jellinek62,liljeblad17,irajizad19,emelyanenko20,ronneberg20,rehfeld24} 

The  stresses obtained from our simulation appear to be consistent with previous computer simulation studies,\cite{xiao16,cui24} but  an order of magnitude smaller than others.\cite{ronneberg20,sun22}  However, despite the qualitative agreement with experiment as regards the role of hydrophilicity and the difference between shear and tensile tests,
the adhesion strengths from simulations are at least two orders of magnitude larger than those measured in laboratory experiments.\cite{jellinek62,liljeblad17,irajizad19,emelyanenko20,ronneberg20,rehfeld24} 

We believe this can be related to two important effects. 

First, our results are performed over perfectly smooth atomic substrates. On the contrary, usual experiments are performed on real world materials which exhibit roughness at scales well beyond the atomic scale. As an example, well controlled experiments are performed only after polishing for matters of standarization, but the polishing produces 'smooth' materials with roughness in the scale of hundreds of nanometers.\cite{rehfeld24} This is well beyond the scale that can be probed in computer simulations, which are usually atomically smooth (c.f. Ref.\cite{cui24} for an exception). The difference in adhesion strength is then likely in part attributed to the real ice-substrate contact area. We speculate that   our simulations provide adhesion strength figures for true ice-substrate contact areas, and the experimental adhesion strengths are smaller because the ice-substrate contact areas are much smaller than the nominal surface area of the interface. This difficulty highlights that adhesion is a problem involving a wide range of length scales. This concept is regularly used in the study of ice friction,\cite{tusima11} and quite generally poses a formidable challenge in the study of tribology.\cite{muser17}

Secondly, our results are performed for ice in contact with the pristine material. On the contrary, usual materials are strongly contaminated by adsorbed impurities, mostly, organic molecules. We speculate that adsorption of such molecules decreases the true contact area beyond that expected from the actual roughness. Indeed, experiments show that the contact angle of water molecules on well controlled clean surfaces increases over time as organic impurities from the atmosphere gradually adsorb onto the surface.\cite{schrader84,osman96,korczeniewski21,orejon24} The same effect, regarding ice is expected to decrease significantly the contact area.

\subsection{4.5. Ice nucleation ability}

Our results provide direct information on the propensity of liquid water to spread in between the interface of ice with a substrate. For our choice of model substrate made of neutral atoms with tunable hydrophilicity, we see that ice has a clear propensity to interfacially premelt even for the most hydrophobic of the substrates studied (including the limit of extreme hydropobicity corresponding to ice in contact with vapor). In practice, systems with contact angles smaller than 50$^\circ$ do not strictly wet the ice/substrate interface, but they are indeed very close to a wetting state and form quasi-liquid layers of one nanometer thickness in the proximity of the melting point. In wetting theory, such propensity may be monitored by the study of the contact angle of a water droplet formed at the interface of ice with the substrate, namely, $\cos\theta_{Swi}=(\gamma_{Si}-\gamma_{Sw})/\gamma_{iw}$, which, according to Eq.\ref{eq:gmin2} may also be given as 
$\cos\theta_{Swi}=g(h_e)/\gamma_{iw} + 1$, readily indicating that water will completely wet the ice/substrate interface for $g(h_e)\to 0$. 

Interestingly, this information is fully complementary to the study of ice nucleation ability, which can be studied rigorously by the calculation of the related phenomena of ice spreading in between the  interface of water with a substrate.\cite{marks23,vicars24} In that case, the relevant parameter is the contact angle formed by a frozen ice parcel formed in between the substrate and water, as dictated by $\cos\theta_{Siw}=(\gamma_{Sw}-\gamma_{Si})/\gamma_{iw}$. This shows that within this  framework of asumed hemispherical droplets (possibly arguable under some circunmstances) it must hold that $\cos\theta_{Siw}=-\cos\theta_{Swi}$. Therefore, our study on interfacial premelting provides interesting information on the nucleation ability of a substrate which is fully complementary  to  the direct calculation of  $S_{Siw}$ by thermodynamic integration \cite{marks23,vicars24} 

Since, as we can see from Table~\ref{tab:hamaker-fit}, $\cos\theta_{Swi}$ is very nearly equal to unity for all substrates, it follows that the complementary $\cos\theta_{Siw}$ must be close to $-1$, and therefore, all the surfaces studied here must be extremely reluctant to act as heterogeneous drivers of ice formation. Furthermore, in the special case that liquid droplets freeze at the air/substrate interface, a quasi-liquid layer of water will appear in between ice and the substrate, facilitating the removal of ice by use of shear forces. 

At any rate, we can notice from Table~\ref{tab:hamaker-fit}, that very hydrophilic substrates are unlikely to act as good nucleators, at least in the case of apolar surfaces. Firstly, because the more hydrophilic a substrate is, the more likely it is to exhibit complete interfacial premelting; secondly, because the strong adsorption layer formed right next to the substrate often consumes all hydrogen bonds in a two dimensional network, therefore  inhibiting further ice formation.\cite{fitzner20,li21b,huang24}

Based on this view, the problem of ice nucleation appears as a problem of competing adsorption of different forms of water. Indeed, colloidal models with interactions that enforce tetrahedrality, as in water, but that do not have attractions of sufficient range to promote condensation, are able to nucleate and spread on smooth substrates.\cite{baran23} This suggests that deposition nucleation from the vapor phase could possibly occur on this kind of substrates provided the thermodynamic conditions are sufficiently far from the liquid-vapor condensation line. Therefore, 
it would be instructive to test the substrates studied herein at the 
conditions where ice could be directly deposited from its vapor
without an intruding liquid adsorption layer,\cite{roudsari24} i.e., at conditions of undersaturation with respect to liquid water.

From this discussion, it appears that at conditions close to liquid-vapor coexistence, the most relevant characteristics of an ice nucleator could be a moderately hydrophilic character, together with a decorated motive of charges or dipoles able to form hydrogen bonds with water.\cite{vicars24,huang24}

\section{5. Conclusion}

In this work we have studied the phenomenon of interfacial premelting of ice  in contact with a model substrate of variable wettability. The model substrate is a smooth closed packed atomic solid of FCC structure with the (111) plane exposed in perfect match with the ice basal plane. This appears to be an optimal situation for the freezing of water on the substrate. However, for substrates of differing wettability, ranging from very hydrophobic ($\theta=120^\circ$) to moderately hydrophilic ($\theta=50^\circ$) we find that an equilibrium quasi-liquid layer is formed spontaneously in all cases. With due reservations related to our choice of atomic and neutral model substrate of either hydrophobic or moderately hydrophilic nature, we find that:
\begin{itemize}
    \item All substrates spontaneously form a quasi-liquid film at temperatures as low as 230~K.
    \item The film grows from about one single disordered water layer at 230~K, to at least one nanometer thick water layer close to the melting point (c.f. Fig~\ref{fig:h_v_T_and_p}). 
     \item  For the most hydrophobic substrate ($\theta=120^\circ$) the structure of interfacially premelted films very much resembles the structure of films formed at the ice/vapor surface (c.f. Fig~\ref{fig:density_p1_f1}).
    \item For the remaining substrates, of higher hydrophilicity, the interfacial films are significantly thicker than the premelting films found at the ice/vapor interface, i.e. the premelting structure at the free ice surface need not be a good proxy for interfacially premelted films formed on a hydrophilic substrate (c.f. Fig~\ref{fig:density_p1_f4}).
    \item Overall we find that the premelting film thickness increases systematically as the substrate becomes more hydrophilic (c.f. Fig~\ref{fig:h_v_T_and_p}).    
\end{itemize}
Our model systems have been described in the light of wetting theory, with the useful tool of interface potentials. From this analysis we show that:
\begin{itemize}
    \item Within a broad range of temperatures and pressures spanning about 20~K and 1000~atm,
    the premelting properties appear to depend on one single thermodynamic variable, namely, the
    pressure difference between bulk ice and water pressure at the corresponding equilibrium ice pressure, $\Delta p$ (c.f. Fig.~\ref{fig:h_v_Pi_all_f}). By this token, we are able to correlate film thicknesses at different thermodynamic states but equal $\Delta p$. The same appears to hold true for the whole interfacial structure as described by the density profiles (c.f.Fig.~\ref{fig:T262@p_400_vs_T260@p1}-\ref{fig:T262@p800_vs_T269@p1}).
    \item The van der Waals forces contribute only a very small fraction of the full surface intermolecular forces. In the range of a few nanometer thickness and below, the structural forces with exponential decay completely dominate (c.f. Fig.~\ref{fig:disjoining-contributions-temp}). The van der Waals forces are only relevant for films more than 1~nm thick. This corresponds to conditions that are attained only within an extremely narrow temperature regime close to melting. A model of exponential forces should be preferred, but van der Waals forces are important to describe the absence of complete surface premelting in moderately hydrophilic to hydrophobic substrates.
    \item We have provided a model of interface potentials featuring short and long range contributions. The model is consistent with the results of computer simulations but allows for extrapolation beyond the conditions probed in simulation (c.f. Eq.\ref{eq:expandvdw}). We expect these results will be useful for the modeling of interfacial premelting in theories of ice adhesion, regelation and friction (c.f. Supplementary Information).
    \item Our model  interface potentials point to absence of complete interfacial melting for substrates that are either hydrophobic or moderately hydrophilic. Under these conditions short range structural forces favor wetting, but long range van der Waals forces inhibit wetting. However, extrapolation of our results show that interfacial premelting films at bulk ice-water coexistence completely wet hydrophilic substrates with contact angle less than $\theta=50^\circ$ (c.f. Table~\ref{tab:hamaker-fit}).
\end{itemize}
The simulation of interfacially premelted systems at negative pressures effectively allows us to perform tensile strength tests. The results can be compared with shear adhesion tests performed earlier.\cite{baran22} As a conclusion we find: (c.f. Fig.\ref{fig:shear-v-tensile})
\begin{itemize}
    \item Adhesion on {\em clean and atomically smooth substrates} increases with the hydrophilicity of the substrate, ranging from less than 100~atm for strongly hydrophobic substrates to more than 2000~atm for hydrophilic substrates.
    \item The shear adhesion strength is similar to the tensile adhesion strength for the most hydrophobic materials, but becomes many times larger for the most hydrophilic materials.
    \item The perpendicular pressure exercised upon shear does not appear to have a large effect on the shear adhesion strength.
    \item Adhesion strengths obtained from simulation are a factor of 100 times larger than usual experimental values. We speculate that this difference is the result of surface roughness and adsorbed impurities in real world materials.
\end{itemize}
Finally, we have explored the wetting properties of interfacially premelted films as a simple and convenient proxy for the understanding of heterogeneous ice nucleation. Our results suggest that neutral substrates of whatever hydrophilicity are very unlikely to act as ice nucleators, with the hydrophobic substrates being more close to promote nucleation than the hydrophilic substrates. For this reason, it seems that good ice nucleators need a combination of moderate hydrophilicity and a decorated motif of charges, dipoles or hydroxyl groups able to match the ice lattice. Otherwise, the large hydrophilicity leads to the promotion of surface melting, and inhibits the possibility of ice nucleation.

\begin{acknowledgement}
LGM  would like to thank financial support from the Agencia Estatal de Investigaci\'on (Ministerio de Ciencia y Econom{\'i}a) under grant PID2023-151751NB-I00. \L B would like to thank the Polish National Agency for Academic Exchange, under Grant No. BPN/BEK/2023/1/00006 Bekker 2023.  We also benefited from generous allocation of computer time at the Academic Supercomputer Center (CI TASK) in Gdansk. 
\end{acknowledgement}

\begin{suppinfo}
    \begin{itemize}
        \item Additional details for the estimation of melting pressure and Hamaker constant calculations. Validation of piston barostat. Tables S1-S6. Figures S1-S3. 
    \end{itemize}
\end{suppinfo}

\begin{tocentry}
    \includegraphics[]{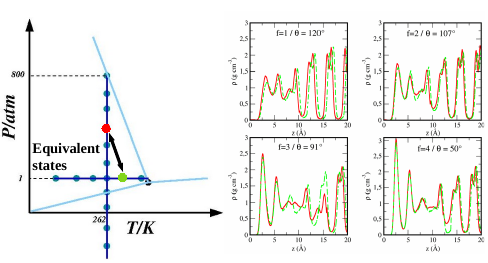}
\end{tocentry}

\bibliography{new}

\clearpage

\includepdf[pages=-]{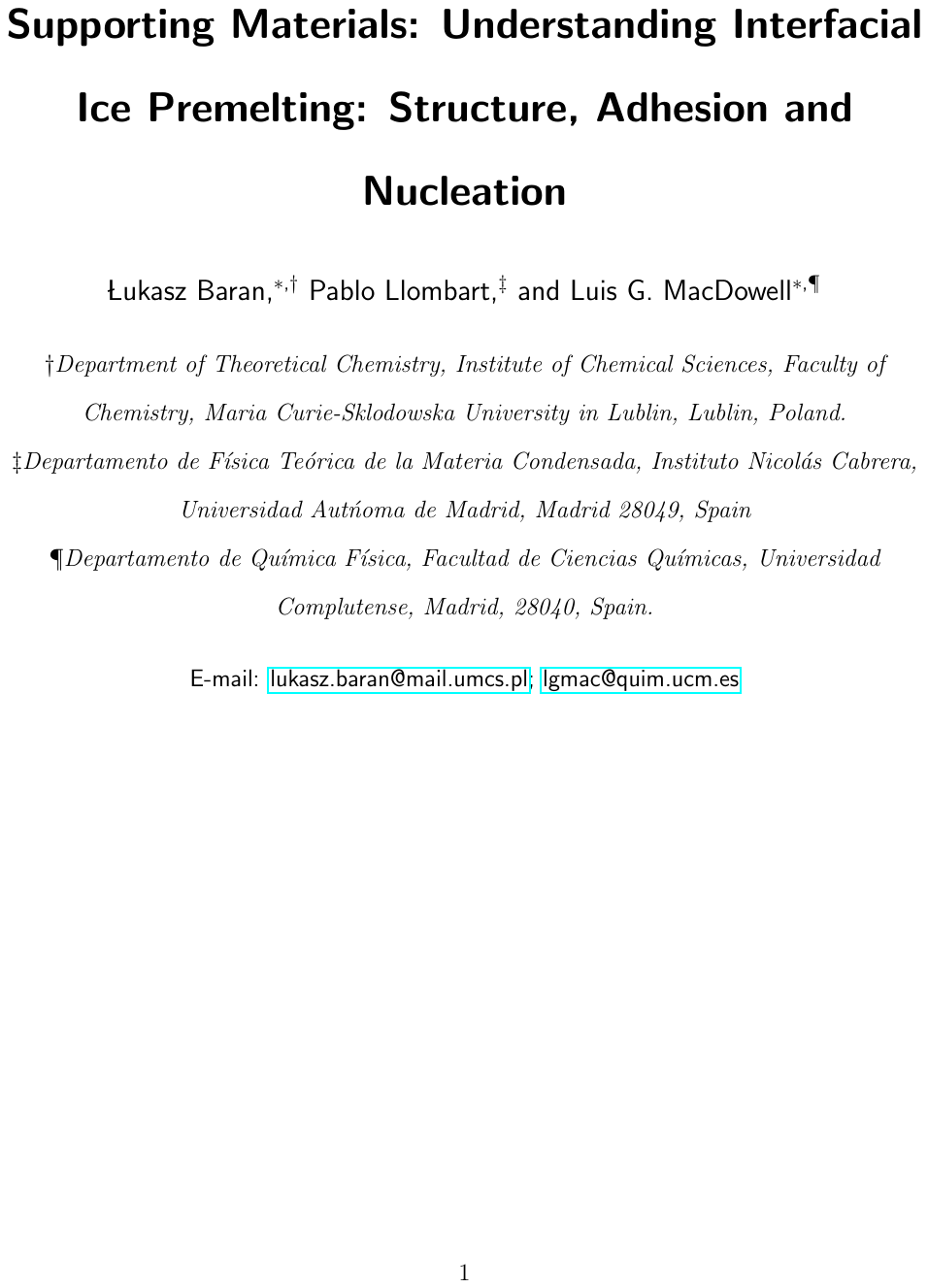}

\end{document}